\documentclass[10pt,twocolumn,letterpaper]{article}

\usepackage[pagenumbers]{cvpr}

\definecolor{cvprblue}{rgb}{0.21,0.49,0.74}
\usepackage[pagebackref,breaklinks,colorlinks,allcolors=cvprblue]{hyperref}

\usepackage{nicefrac}

\title{Bridging the Gap between Gaussian Diffusion Models and Universal Quantization for Image Compression}

\author{
    Lucas Relic$^{1,2}$ \quad
    Roberto Azevedo$^{2}$ \quad
    Yang Zhang$^{2}$ \quad
    Markus Gross$^{1,2}$ \quad
    Christopher Schroers$^{2}$
    \vspace{0.3em} \\
    {$^1$ETH Z\"urich} \quad
    {$^2$DisneyResearch\textbar Studios}
}

\begin{document}
\twocolumn[{%
\renewcommand\twocolumn[1][]{#1}%
\maketitle
\begin{center}
    \centering
    \captionsetup{type=figure}
    \includegraphics[width=.99\textwidth]{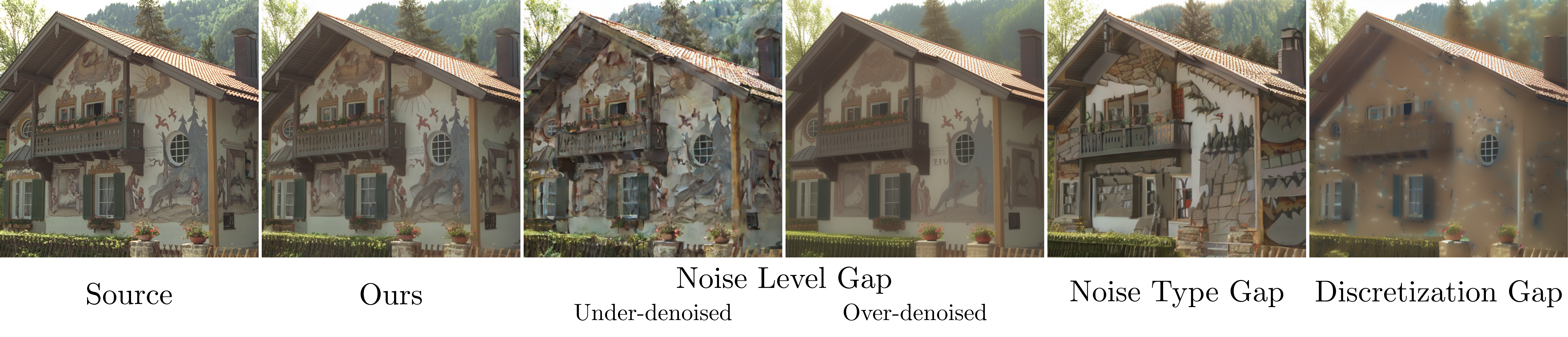}
    \captionof{figure}{Visualization of the 3 gaps we address in this work. Failure to match the noise level~(middle columns) results in either too noisy or too smooth images. Inconsistent noise types~(middle-right) introduces generative artifacts and color shift. Applying diffusion to discrete data (far right) causes flat textures as well as color shift. Addressing all three gaps~(middle-left) results in the most realistic reconstruction that best matches the source image~(far left).}
  \label{fig:snr_mismatch}
\end{center}%
}]


\begin{abstract}
Generative neural image compression supports data representation at extremely low bitrate, synthesizing details at the client and consistently producing highly realistic images.
By leveraging the similarities between quantization error and additive noise, diffusion-based generative image compression codecs can be built using a latent diffusion model to ``denoise'' the artifacts introduced by quantization.
However, we identify three critical gaps in previous approaches following this paradigm~(namely, the noise level, noise type, and discretization gaps) that result in the quantized data falling out of the data distribution known by the diffusion model.
In this work, we propose a novel quantization-based forward diffusion process with theoretical foundations that tackles all three aforementioned gaps.
We achieve this through universal quantization with a carefully tailored quantization schedule and a diffusion model trained with uniform noise.
Compared to previous work, our proposal produces consistently realistic and detailed reconstructions, even at very low bitrates.
In such a regime, we achieve the best rate-distortion-realism performance, outperforming previous related works.
\end{abstract}  


\section{Introduction}
\label{sec:intro}

In today’s data-driven world, the field of neural image compression (NIC)~\cite{balle2017Endtoend, balle2018Variationala, theis2017Lossy} has experienced significant growth, with increasing demand for more effective codecs.
As NIC performance continues to improve, recent efforts have focused on achieving compression at even lower bitrates~\cite{mentzer2020HighFidelitya, relic2024Lossy}.
Here, generative models excel, leveraging their ability to synthesize textures effectively under stringent information constraints.

Recently proposed methods~\cite{hoogeboom2023HighFidelity, yang2023Lossy, relic2024Lossy} use diffusion models~\cite{sohl-dickstein2015Deep,ho2020Denoising} as an expressive decoder to produce highly detailed, realistic reconstructions, especially at extremely low bitrates.
This is achieved by conditioning the diffusion model on information extracted from the source image and generating a new image that attempts to match the source as closely as possible.
We refer to this as the Conditioning-based Diffusion Image Compression~(CDIC) strategy.
However, latent diffusion models~\cite{rombach2022HighResolution} also support a different paradigm for lossy image compression.
Quantization error can be modeled as noise~\cite{gray1998Quantizationa, balle2017Endtoend}, and given that diffusion models are denoising models, one can directly apply them to the data to remove quantization artifacts~\cite{relic2024Lossy}.
We coin this strategy ``Dequantization''-based Diffusion Image Compression~(DDIC).
Following this paradigm provides several benefits, such as reduced decoding time~(due to the small number of diffusion steps compared to the full generation process) and increased flexibility to use foundation models as minimal architecture changes are needed. 

Although DDIC is a promising approach, we identify three gaps in previous works in this area~(detailed in Section~\ref{sec:gaps}): the \emph{noise type}, the \emph{noise level}, and the \emph{discretization} gaps.
The \emph{noise type gap} represents the difference in distribution between quantization error and Gaussian diffusion models.
The \emph{noise level gap} refers to the possible mismatch in the expected signal-to-noise ratio of the partially noisy data versus the actual ratio.
The \emph{discretization gap} arises from passing discrete data to a continuous diffusion model.
Leaving these gaps unsolved causes the data to fall out of the distribution of the diffusion model, negatively impacting the final reconstruction quality~(see Fig.~\ref{fig:snr_mismatch}).

To tackle the above gaps, we propose a new theoretically-founded quantization-based diffusion forward process that places the quantized data perfectly along the diffusion trajectory.
Our proposed forward process uses universal quantization to close the discretization gap and introduces a new quantization schedule that dictates the signal-to-noise ratio of the quantized data -- which closes the noise level gap.
Finally, we solve the noise type gap using a diffusion model trained with uniform noise, thus matching the distribution of the quantization error.
We additionally show that such a uniform noise diffusion model can be efficiently obtained by fine-tuning existing Gaussian diffusion models.
Following our proposal, we build an image codec that produces more realistic and detailed reconstructions than previous methods while being able to operate at a wider range of target bitrates.

In summary, our contributions are:
\begin{itemize}
    \item We identify three gaps that negatively affect the performance of DDIC codecs: i)~the noise type gap, ii)~the noise level gap, and iii)~the discretization gap.
    \item We propose a novel diffusion-based image codec which solves all the three gaps.
    We introduce a novel quantization-based forward diffusion process, to close the discretization and noise level gaps, and utilize a uniform noise diffusion model to close the noise type gap.
    \item We establish the validity of latent uniform noise diffusion models and show that one can be efficiently obtained by finetuning a foundation Gaussian diffusion model.
    \item We validate our proposed method on various datasets and evaluation criteria, showing improved quantitative and qualitative results, particularly at very low bitrates.
\end{itemize}

\vspace{-0.25em}


\section{Background and Related Work}
\label{sec:rel_work}

\subsection{Neural Image Compression}
Neural image compression~(NIC) codecs~\cite{balle2017Endtoend, balle2018Variationala, minnen2018Jointa, mentzer2020HighFidelitya} convert between images and bitstreams via transform coding~\cite{balle2017Endtoend}, where an image \(\mathbf{x}\) is transformed to a representation \(\mathbf{y}\) that is then converted to bitstream.
In such methods, the forward transform \(g_a\) and reverse transform \(g_s\) are parameterized by a neural network of arbitrary architecture, \eg, a VAE~(Variational Autoencoder)~\cite{kingma2022AutoEncoding} or a GAN~(Generative Adversarial Network)~\cite{goodfellow2014Generative}.
Converting between latent and bitstream, performed by an entropy model~\cite{balle2017Endtoend, balle2018Variationala, minnen2018Jointa}, is similarly parameterized by a neural network.
Critically, encoding to bitstream requires discrete symbols, and thus, the continuous output of \(g_a\) must first be discretized.
Formally,
\begin{equation}
\label{eq:nic}
     \hat{\mathbf{y}} = \lfloor g_a(\mathbf{x}) \rceil,\quad\hat{\mathbf{x}} = g_s(\hat{\mathbf{y}}),
\end{equation}
where \(\lfloor \cdot \rceil\) denotes the rounding operation, \(\hat{\mathbf{y}}\) is the quantized latent representation, and \(\hat{\mathbf{x}}\) is the reconstructed image.
This discretization results in a loss of information and introduces error, negatively affecting the reconstruction quality.
A common solution is to simulate this behavior during optimization~\cite{balle2017Endtoend, balle2018Variationala, agustsson2020Universally}, training the networks to be robust to the error despite the discrepancy between the encoded discrete representation and continuous data seen during training.

One advantage of NIC compared to traditional~(non-learned) compression methods is that they can be directly optimized for the rate-distortion tradeoff, \ie, the balance between compression cost and reconstruction quality:
\begin{equation}
\label{eq:rd}
    \mathcal{L} = R(\hat{\mathbf{y}}) + \lambda \cdot D(\mathbf{x}, \hat{\mathbf{x}}),
\end{equation}
where \(R(\hat{\mathbf{y}})\) is the cost of compression, \(D(\mathbf{x}, \hat{\mathbf{x}})\) is a metric representing the similarity between the input and output images, and \(\lambda\) is a hyperparameter controlling the tradeoff between the two terms.

\subsection{Diffusion Models}
Diffusion models~\cite{sohl-dickstein2015Deep, song2021Denoising} define a process that models the transition between random noise and structured data.
When the forward~(data to noise) and reverse~(noise to data) processes are divided into small steps, the transition between each step is the addition or removal of a Gaussian noise sample. The full diffusion process is thus a traversal between a series of timesteps \(t\in[N,0]\).
While this process is iterative, one can also express the partially noisy diffusion variable \(\mathbf{y}_t\) at any given \(t\) in terms of the original data \(\mathbf{y}_0\) and a noise sample \(\epsilon\):

\begin{equation}
\label{eq:diffusion_forward}
    \mathbf{y}_{t} = \sqrt{\alpha}_{t} \mathbf{y}_{0} + \sqrt{1 - \alpha}_{t}\epsilon, \quad \epsilon \sim \mathcal{N}(0,\mathbf{I}),
\end{equation}

where \(\alpha_{t}\)~(known as the ``variance schedule'') defines the ratio of signal and noise at every \(t\) and increases as \(t\rightarrow0\).
In practice, the reverse diffusion process is intractable and thus parameterized by the diffusion model, which learns to iteratively denoise \(\mathbf{y}_t\) by stepping through \(t=\{N,...,1,0\}\).
In DDIM~\cite{song2021Denoising}~(the sampling strategy of diffusion models we use in this work) the partially denoised data \(\mathbf{y}_{t-1}\) can be computed from the noisy data \(\mathbf{y}_{t}\) with:
\begin{equation}
\label{eq:ddim}
    \mathbf{y}_{t-1}=\sqrt{\alpha_{t-1}}\tilde{\mathbf{y}}_0+\sqrt{1-\alpha_{t-1}}\epsilon_\theta(\mathbf{y}_t, t)
\end{equation}
where \(\epsilon_\theta(\mathbf{y}_t, t)\) is a forward pass of the diffusion model, which takes \(\mathbf{y}_t\) and the current timestep \(t\) as input, and \(\tilde{\mathbf{y}}_0 = f(\epsilon_\theta(\mathbf{y}_t, t), t)\) is an estimation of the fully denoised data which is computed from the output of the diffusion model and the current timestep.

Latent diffusion models~\cite{rombach2022HighResolution} move the diffusion process to the latent space of a VAE.
This yields efficiency gains as the diffusion model operates in a lower-dimensional representational space, allowing for image generation at high resolutions.
Our proposed pipeline~(Section~\ref{sec:method}) builds on latent diffusion, taking advantage of such efficiency gains for generative image compression.

\subsection{Diffusion-based Image Compression}
Most diffusion-based image compression works follow the CDIC paradigm, in which an efficient data representation is extracted from the image and then used to condition the generative diffusion process at the decoding time.
The conditioning signal often takes the form of some spatial information, such as a learned embedding~\cite{yang2023Lossy, careil2023image}, an image compressed via another codec~\cite{hoogeboom2023HighFidelity, goose2023Neural}, or an edge or color map~\cite{lei2023Text, bachard2024CoCliCo}.
It can also be an unstructured content variable, for example, text from an image captioning model~\cite{careil2023image} or a CLIP embedding~\cite{lei2023Text, bachard2024CoCliCo}.
The extracted conditioning signal is then injected into the diffusion generation process by concatenation to the diffusion model input~\cite{hoogeboom2023HighFidelity, goose2023Neural, careil2023image} or intermediate layers~\cite{yang2023Lossy}, via cross-attention~\cite{lei2023Text, careil2023image, bachard2024CoCliCo, pan2022Extreme}, or through a ControlNet~\cite{zhang2023Addinga, lei2023Text}.
Regardless of the modality, such a paradigm often requires training the diffusion model from scratch so that it can accept the respective conditioning modality.
Additionally, due to the iterative diffusion process, 
conditionally sampling an image requires a long decoding time, limiting practicality.

Most similar to our proposal, Relic et al.~\cite{relic2024Lossy} follow the dequantization paradigm~(DDIC) and use a latent diffusion model to remove artifacts introduced during quantization.
They adaptively quantize the data by training a parameter estimation module that predicts quantization parameters, introducing a variable amount of error to the signal.
Therefore, they must predict the number of denoising steps to perform at the receiver, corresponding with the quantization noise.
While following the dequantization paradigm brings them substantial gains in computational efficiency, Relic~\etal's formulation suffers in the three areas discussed in Sec.~\ref{sec:gaps}, which results in sub-optimal reconstructions and a limited range of practical target bitrate.
As will be clear during the next two sections, in this work we solve such issues proposing a method that has the advantages of Relic~\etal, while still allowing realistic constructions on a broader range of bitrates.

Finally, like our proposal, a concurrent work~\cite{2024Progressive} also investigates uniform noise diffusion models in the context of image compression.
They formulate a diffusion model using uniform noise whose objective function corresponds to compression cost and build a progressive codec around this model.
While encouraging, their method focuses on bitrates several orders of magnitude larger than our proposal.
Additionally, it operates in the computationally expensive pixel domain and thus their method is only shown to be effective on small resolution images.
As a result, the practical use of their codec is not yet feasible.
In contrast, by building on the latent diffusion framework, our proposal is significantly more efficient than~\cite{2024Progressive} and is able to work with high-resolution images.


\section{Open problems in DDIC Methods}
\label{sec:gaps}

\subsection{Noise Type Gap}
Quantization error in many domains~(such as the latent domain) is commonly known in signal processing to be well approximated by \emph{uniform} noise~\cite{gray1998Quantizationa, balle2017Endtoend}.
However, diffusion models assume a \emph{Gaussian} noise structure as it aligns with natural data distribution assumptions and facilitates tractable modeling.
This results in the \emph{noise type gap} ---a discrepancy between the quantization error~(well approximated by uniform noise) and the Gaussian noise used in the diffusion process.
This misalignment means that when a Gaussian denoising model interacts with uniform quantization noise, the model fails to correctly predict the actual noise characteristics, resulting in generative artifacts. (see Fig.~\ref{fig:snr_mismatch}, \emph{Noise Type Gap}).
Specifically, the mismatch can lead to visually disruptive effects such as unnatural color shifts, texture inconsistencies, and artificial patterns that degrade the realism and fidelity of the generated image.

While theoretical frameworks exist which support uniform noise diffusion models~\cite{heitz2023Iterative, peluchetti2023NonDenoising, 2024Progressive} they are shown to be effective only on toy datasets and small resolution images.
This limitation prevents the real-world applicability of uniform diffusion models, and thus any practical codec following the dequantization paradigm must use a Gaussian diffusion model and suffers from the noise type gap.

\begin{figure*}[ht!]
  \centering
  \includegraphics[width=\linewidth]{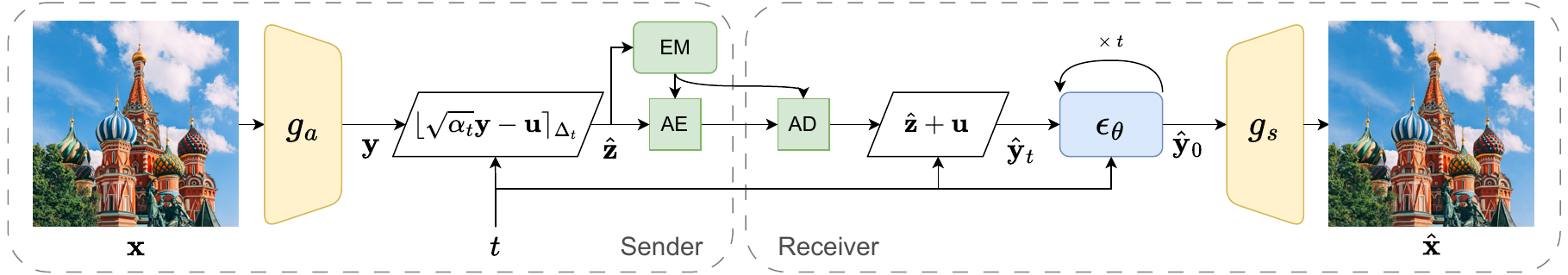}
  \caption{Architecture of our proposed method. An input image is first encoded to the latent space of a diffusion model and discretized according to the quantization stage of our proposed forward process. The discrete data is transmitted across the channel, subject to the post-quantization stage of our forward process, denoised by the diffusion model, and decoded back to image space. The user-input timestep parameter dictates the quantization parameters according to our proposed quantization schedule, as well as the number of denoising steps performed by the diffusion model.}
  \label{fig:method}
\end{figure*}

\subsection{Discretization Gap}
The neural decoders in NIC methods, despite being continous models, must operate on discrete representations extracted from the transmitted bitstream;
most methods build robust decoders which minimize the resulting negative effects~\cite{balle2017Endtoend, balle2018Variationala}.
However, building a similarly robust diffusion model in this context is impossible, since they model transitions between continuous states and are inherently unable to handle discrete inputs.
We term this behavior the \emph{discretization gap} ---the incompatibility between using discrete input data with continuous diffusion models.\footnote{Note our definition represents a different, although related, phenomenon than described in Yang~\etal~\cite{yang2020Improving}, despite the same name.}
Under the discretization gap, small variation in the input data is eliminated, which leads to flat textures and loss of detail, and using a large quantization bin size causes blocking artifacts and color shifts due to the low resolution of the color palette~(Fig.~\ref{fig:snr_mismatch}, \emph{Discretization Gap}).

\subsection{Noise Level Gap}
Diffusion image generation assumes a fixed progression through the variance schedule, which dictates the noise level at each \(t\).
It is therefore critical to ensure a match in noise level between the forward and backward process~(\ie, \(t\) must be the same in both Eq.~\eqref{eq:diffusion_forward} and Eq.~\eqref{eq:ddim});
failure to do so violates the equations which form the theoretical basis of diffusion models.
However, when using a different forward process, as done in DDIC, it is possible for the forward and reverse processes to not align.
This is the \emph{noise level gap} - a difference in the actual noise level of the diffusion variable versus what is expected at any timestep.
Intuitively, the diffusion model either over- or under-estimates the noise in the variable throughout the diffusion process, which results in either noisy or overly smoothed image reconstructions~(Fig.~\ref{fig:snr_mismatch}, \emph{Noise Level Gap)}.

Other works which follow the DDIC paradigm~\cite{relic2024Lossy} estimate both quantization parameters~(\ie how much noise is introduced) and the number of denoising iterations~(\ie how much noise to remove), thus estimating independent \(t\)s for Eqs.~\eqref{eq:diffusion_forward} and~\eqref{eq:ddim}.
While this technique may predict approximately close \(t\)s, small changes in the variance schedule have significant impact on final image quality~\cite{hoogeboom2023simplea, chen2023Importance}.
Thus, even a well learned approximation of noise level produces suboptimal end results and care should be taken to eliminate the noise level gap completely.


\section{Method}
\label{sec:method}

Next, we build a solution that solves the three gaps identified in Section~\ref{sec:gaps}.
Our pipeline, shown in~\cref{fig:method}, follows the DDIC paradigm of a latent diffusion model with our proposed forward process.

\subsection{Universal quantization diffusion compression}
\label{sec:meat}

\paragraph{Improved forward diffusion process}
One of our key contributions to closing the aforementioned gaps is an improved forward diffusion process.
The goal is to formulate this process with quantization, such that a discrete variable can be encoded to bitstream, while maintaining the noise characteristics of the original diffusion variance schedule.
We begin with the standard forward process~(a slight reorganisation of Eq.~\ref{eq:diffusion_forward}):
\begin{equation}
\label{eq:diffusion_smalleps}
    \mathbf{y}_{t} = \sqrt{\alpha}_{t} \mathbf{y}_{0} + \epsilon, \quad \epsilon \sim \mathcal{N}(0, (1 - \alpha_{t})\mathbf{I}),
\end{equation}

We propose to introduce universal quantization to the forward noising process in order to obtain a discrete variable for entropy coding.
Universal quantization~\cite{ziv1985universal, zamir1992universal} is hard quantization dithered by a uniform random variable.
This has the unique property of being equal in distribution to simply adding another sample (from an identical random variable) to the original unquantized variable: 

\begin{equation}
\label{eq:universal_quant}
    \hat{\mathbf{y}} = \lfloor\mathbf{y} - \mathbf{u}\rceil_{\Delta} + \mathbf{u} \stackrel{d}{=} \mathbf{y} + \mathbf{u}', \quad \mathbf{u}, \mathbf{u'} \sim \mathcal{U}[\nicefrac{-\Delta}{2}, \nicefrac{\Delta}{2}],
\end{equation}

where \( \lfloor \cdot \rceil_{\Delta}\) denotes rounding to a bin of width \( \Delta \).

We begin to construct our new forward process by combining Eqs.~\eqref{eq:diffusion_smalleps} and~\eqref{eq:universal_quant} and separating into quantization and post-quantization stages:
\footnote{Note \(\Delta\) and \(\Delta_t\) are equivalent, we use the latter to explicitly denote that \(\Delta\) can vary as a function of the diffusion step \(t\).}

\begin{gather}
\label{eq:uddq}
\hat{\mathbf{y}}_t = \lfloor \sqrt{\alpha}_{t}\mathbf{y} - \mathbf{u} \rceil_{\Delta_t} + \mathbf{u}, \quad \mathbf{u} \sim \mathcal{U}[\nicefrac{-\Delta_t}{2}, \nicefrac{\Delta_t}{2}] \\
\label{eq:uddq_split}
\hat{\mathbf{z}} = \lfloor \sqrt{\alpha}_{t}\mathbf{y} - \mathbf{u} \rceil_{\Delta_t}, \quad \hat{\mathbf{y}}_t = \hat{\mathbf{z}} + \mathbf{u}.
\end{gather}

Eq.~\eqref{eq:uddq} already solves the discretization gap.
Compared to hard quantization, which passes the discrete data directly to the decoder~(as in Eq.~\eqref{eq:nic}), our output \(\hat{\mathbf{y}}_t\) is once again a continuous variable;
the addition of a uniform noise sample moves the data back into continuous space.

We now shift our focus on bridging the noise level gap, which is done by matching the signal-to-noise ratio~(SNR) of \(\mathbf{y}_t\) and \(\hat{\mathbf{y}}_t\) for all \(t\).
Via Eq.~\eqref{eq:uddq}, \(\text{SNR}(\hat{\mathbf{y}}_t)\) can be controlled by adjusting the quantization bin width and uniform noise support, defined in terms of \(\Delta_t\).
Thus, to close the noise level gap, we must match the noise levels of \(\mathbf{y}_t\) and \(\hat{\mathbf{y}}_t\):

\begin{equation}
    \text{SNR}(\hat{\mathbf{y}}_t) = \text{SNR}(\mathbf{y}_t) \quad \forall t \in \{T, ..., 0\}.
\label{eq:quant_schedule}
\end{equation}

We therefore introduce a \emph{quantization schedule}, which varies \(\Delta_t\) as a function of \(t\).
Our quantization schedule can be matched with the diffusion variance schedule by substituting Eqs.~\eqref{eq:diffusion_forward} and~\eqref{eq:uddq} into Eq.~\eqref{eq:quant_schedule} and solving for \(\Delta_t\): \footnote{We provide a derivation in the Supplementary Material.}

\begin{equation}
    \Delta_t=\sqrt{12(1-\alpha_t)}
\label{eq:delta_t}
\end{equation}

Our proposed quantization-based forward process simultaneously eliminates both the discretization and noise level gaps, via universal quantization and the quantization schedule, respectively.
An added benefit of our quantization schedule is that \(t\) also becomes a rate-distortion tradeoff parameter, as the quantization bin width directly impacts the final size of the compressed bitstream.
Additionally, because diffusion models can denoise data at any arbitrary timestep, our method supports compression to multiple bitrates with a single model by accepting \(t\) as user input at inference time.

For any \(t=\tau\), \(\Delta_\tau\) is computed via \cref{eq:delta_t} by indexing into \(\alpha_t\) at timestep \(\tau\) and used in \cref{eq:uddq} to produce \(\hat{\mathbf{y}}_{\tau}\), which is denoised by the diffusion model over \(t\in \{\tau,...,1,0\}\).
\(\tau\) must be known by both sender and receiver and is transmitted as side information for negligible bit cost.

\paragraph{Uniform noise diffusion model}

Depsite existing formulations for a uniform noise diffusion model~\cite{2024Progressive, heitz2023Iterative, peluchetti2023NonDenoising}, we have experimentally been unable to train such a model on images of practical resolution with reasonable compute budget.
However,~\cite{2024Progressive} show that uniform diffusion models are equivalent to Gaussian diffusion models as \(t\rightarrow\infty\), and thus we theorize that a uniform diffusion model can be efficiently obtained by starting from a pretrained Gaussian diffusion model.
We find that this can be achieved by finetuning a foundation diffusion model and exchanging the Gaussian noise for uniform noise.
As diffusion models are sensitive to changes in the variance schedule~\cite{chen2023Importance, hoogeboom2023simplea, hoogeboom2023HighFidelity}, we find it most effective to leave it unchanged, despite the change in distribution.
In our scenario of adapting a Gaussian diffusion model to uniform noise, this is done by drawing \( \epsilon \sim \mathcal{U}(-\sqrt{3}, \sqrt{3}) \) in Eq.~\eqref{eq:diffusion_forward} during training.

\subsection{Implementation}

\paragraph{Architecture}
For the architecture of the latent transforms and diffusion model, we use Stable Diffusion v2.1~\cite{rombach2022HighResolution}.
\(g_a\) and \(g_s\) remain frozen and we finetune the diffusion model.
However, we note that our proposed forward process and uniform noise finetuning can be implemented with any pretrained latent diffusion model, and we select Stable Diffusion due to its publicly available code and model weights and its widespread use in the image generation community.

Entropy coding is performed with a mean-scale hyperprior entropy model~\cite{minnen2018Jointa} and asymmetric numeral systems~(both implemented via CompressAI~\cite{begaint2020CompressAI}).
As we take the Stable Diffusion VAE as \(g_a\) and \(g_s\), we utilize only the entropy model, consisting of hyperprior encoder and decoder and conditonal entropy bottleneck.
Due to the significantly fewer latent channels of the Stable Diffusion VAE compared to other VAE-based NIC codecs, we similarly reduce the channel depth of the hyperprior transforms to 32~(\ie, \(M=32\) in~\cite{begaint2020CompressAI}).

\paragraph{Optimization}
Our method is optimized in two distinct stages.
In the first stage, we finetune the pretrained Gaussian diffusion model to operate with uniform noise.
We follow the original training strategy~(including objective function) of Stable Diffusion, except for sampling \(\epsilon\) from a uniform distribution of unit variance rather than a standard normal Gaussian~(see~\cref{sec:meat}).
Our model is trained for 100k steps on a subset of the LAION Improved Aesthetics 6.5+ dataset with batch size 8 and learning rate \(1e^{-5}\).
We additionally employ input perturbation~\cite{ning2023Input} as this has been shown to improve sampling quality.
The VAE encoder and decoder are kept frozen.

In the second stage we freeze \(g_a\), \(g_s\), and the diffusion model and train only the entropy model to efficiently encode \(\hat{\mathbf{z}}\) to bitstream and back.
Notably, since all image transform modules are frozen and the entropy coding stage is lossless, we optimize only on the rate objective of~\cref{eq:rd}.
We randomly sample \(t\in\{1, 5, 10, 20, 30, 40, 45\}\) during training - the function of the entropy model is an accurate probabilty model of the quantized data, and as the distribution of \(\hat{\mathbf{z}}\) is dependent on input parameter \(t\), we vary it to reflect operation conditions at inference time.
This range of \(t\) was chosen as we employ DDIM sampling with a maximum of 50 steps and thus elect to sample a wide range of possible values.
We perform the second training stage over the Vimeo90k dataset and and crop each image to \(256^2\) resolution. We use batch size 8 and learning rate \(1e^{-4}\).


\section{Experiments}
\label{sec:experiments}

\begin{figure*}[ht!]
\centering
    \begin{subfigure}{0.45\textwidth}
        \centering
        \includegraphics[width=\textwidth]{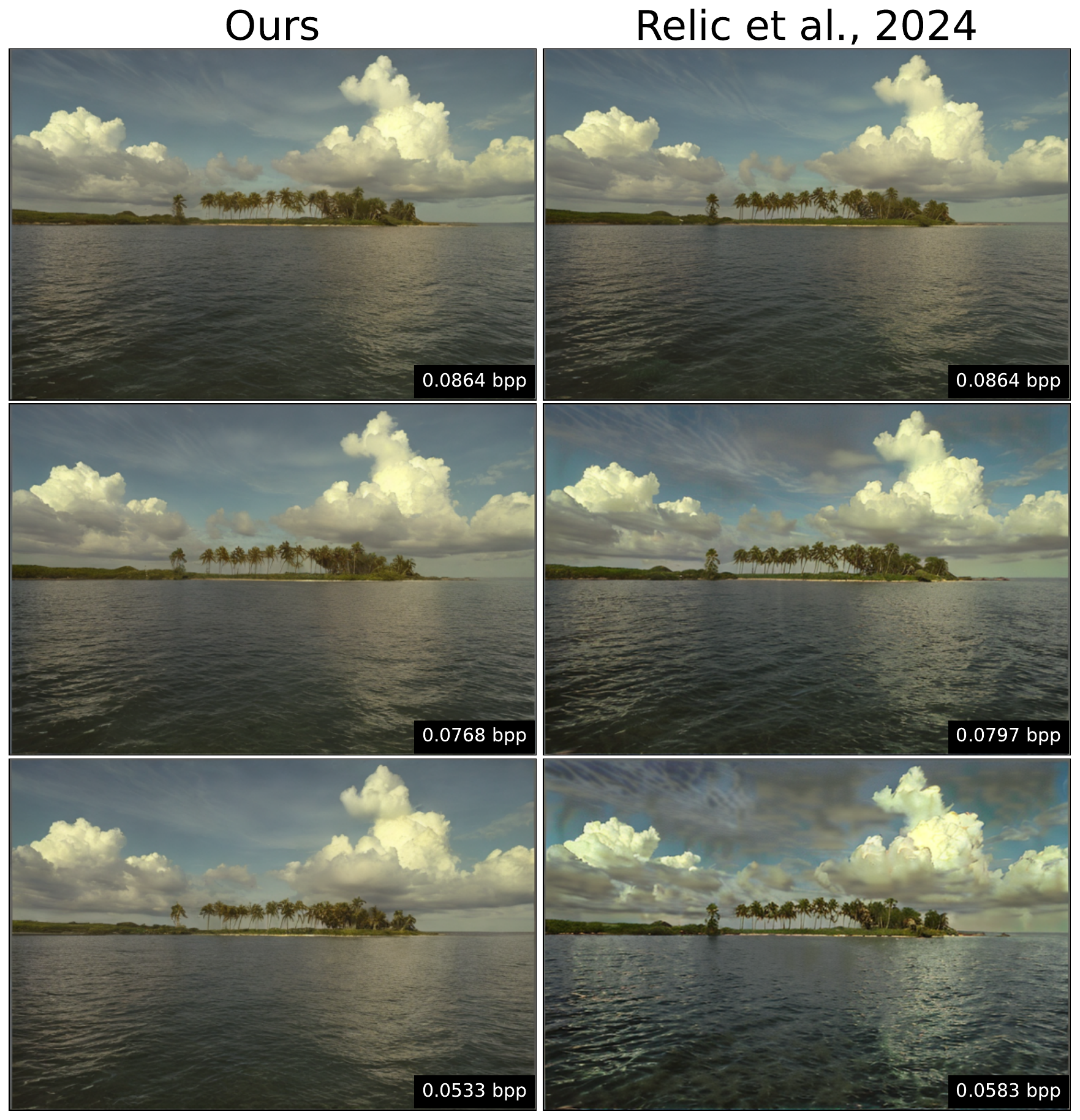}
    \end{subfigure}
    \begin{subfigure}{0.45\textwidth}
        \centering
        \includegraphics[width=\textwidth]{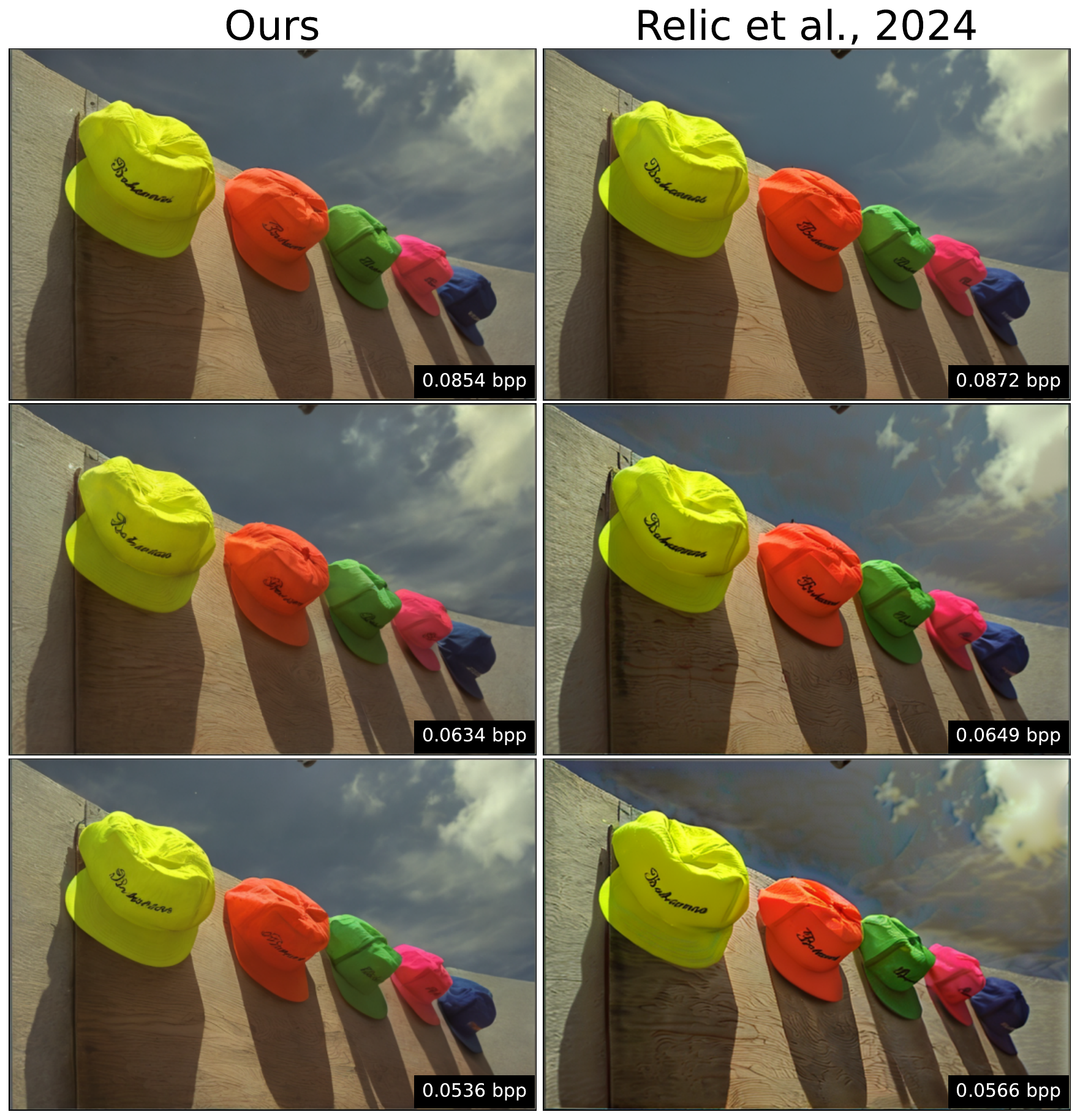}
    \end{subfigure}
    \caption{Progression of image quality between our method and Relic~\etal~\cite{relic2024Lossy} as bitrate decreases. Their reconstructions are significantly worse at lower bitrates while ours maintain high realism at all bitrates. Best viewed digitally.}
\label{fig:bpp-progression}
\end{figure*}

\subsection{Datasets}
We evaluate our method on the following datasets:
\textbf{Kodak}~\cite{kodak1993PhotoCD}, containing 24 images of \(768\times512\) (or transpose);
for larger images, \textbf{CLIC2020}~\cite{Workshop}, containing 428 images approximately 2000px on the longer side;
and \textbf{MS-COCO 30k}, used in recent compression works to measure realism~\cite{agustsson2023MultiRealisma, hoogeboom2023HighFidelity}. We prepare the data as discussed in Agustsson~\etal~\cite{agustsson2023MultiRealisma} to produce 30,000 images of \(256^2\) resolution.

\subsection{Metrics}
For image quality evaluation, we use three primary metrics: FID, LPIPS, and MS-SSIM.
\textbf{FID}~\cite{Heusel2017GANSa} is a metric which assesses the realism of generated images.
We follow recent work~\cite{mentzer2020HighFidelitya, hoogeboom2023HighFidelity, rombach2022HighResolution} and evaluate FID on \(256^2\) image patches.\footnote{The patching process is detailed in Appendix A.7 of Mentzer~\etal~\cite{mentzer2020HighFidelitya}.}
Kodak does not contain enough images to compute FID score, thus on this dataset we evaluate only on the other metrics.
\textbf{LPIPS} serves as a perceptual distortion metric, aiming to assess human-like visual similarity between images.
For pixelwise fidelity, we use \textbf{MS-SSIM}.
Together, these metrics provide a robust evaluation of realism, perceptual similarity, and pixel-level accuracy.

It is important to note that at low bitrates, pixelwise metrics such as MS-SSIM and PSNR do not accurately reflect the quality of reconstructed images~\cite{blau2019Rethinking, careil2023image}.
In fact, it is mathematically proven that achieving high performance in pixelwise distortion necessarily decreases visual quality~\cite{blau2019Rethinking}.
As one of our goals is to produce realistic and perceptually pleasing images, we do not focus on performance measured by MS-SSIM.

\subsection{Baselines}
The codecs proposed by \textbf{Relic~\etal}~\cite{relic2024Lossy}, Hoogeboom/etal~(\textbf{HFD})~\cite{hoogeboom2023HighFidelity}, and Yang and Mandt~(\textbf{CDC})~\cite{yang2023Lossy} are used as baselines to compare against diffusion-based methods.
We only include quantitative comparisons against CDC as it operates in a significantly higher bitrate range, and thus a fair qualitative comparison is not feasible.
Additionally, evaluation with Relic~\etal on CLIC2020 is not possible as their method cannot compress large resolution images.
While other diffusion-based compression methods exist~\cite{careil2023image, bachard2024CoCliCo}, we cannot evaluate on them as no code or reconstructions are available.
We also compare to \textbf{MR}~\cite{agustsson2023MultiRealisma}, \textbf{MS-ILLM}~\cite{muckley2023Improving}, and \textbf{HiFiC}~\cite{mentzer2020HighFidelitya}, all GAN-based approaches, to provide a baseline to other generative image compression codecs.
As a reference of traditional compression methods, take \textbf{VTM} 19.2~\cite{VTM192}, the state of the art in non-learned codecs.

Additional detail on how we produce baseline results is provided in the Supplementary Material.

\subsection{Results}

\begin{figure*}[]
  \centering
  \includegraphics[width=\linewidth]{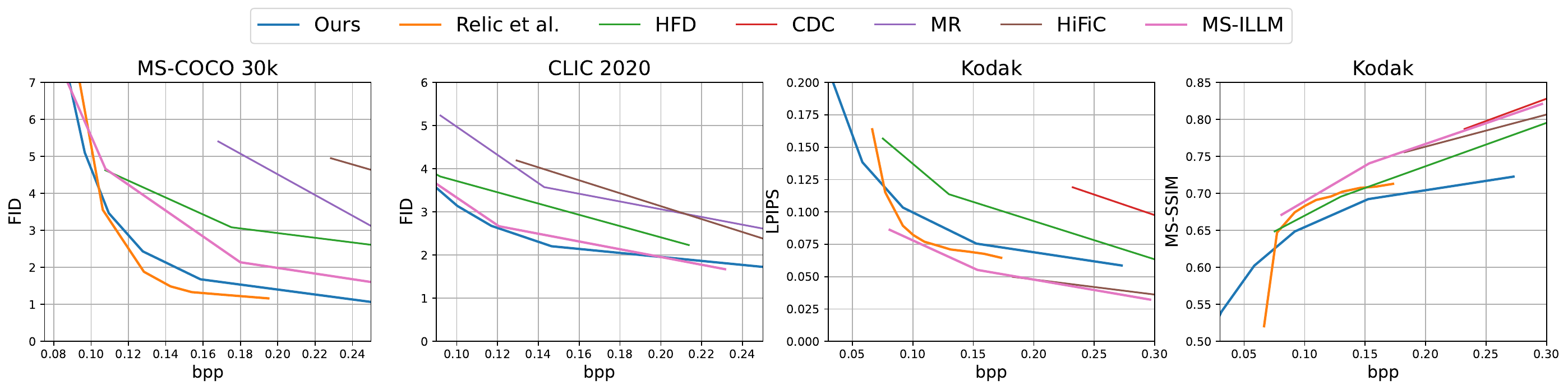}
  \caption{Rate-realism~(left) and rate-distortion~(right) performance of our method compared to Relic et al.~\cite{relic2024Lossy}, HFD~\cite{hoogeboom2023HighFidelity}, CDC~\cite{yang2023Lossy}, MR~\cite{agustsson2023MultiRealisma}, MS-ILLM~\cite{muckley2023Improving}, and HiFiC~\cite{mentzer2020HighFidelitya}.}
  \label{fig:rd}
\end{figure*}

\begin{figure}
    \centering
    \includegraphics[width=0.8\linewidth]{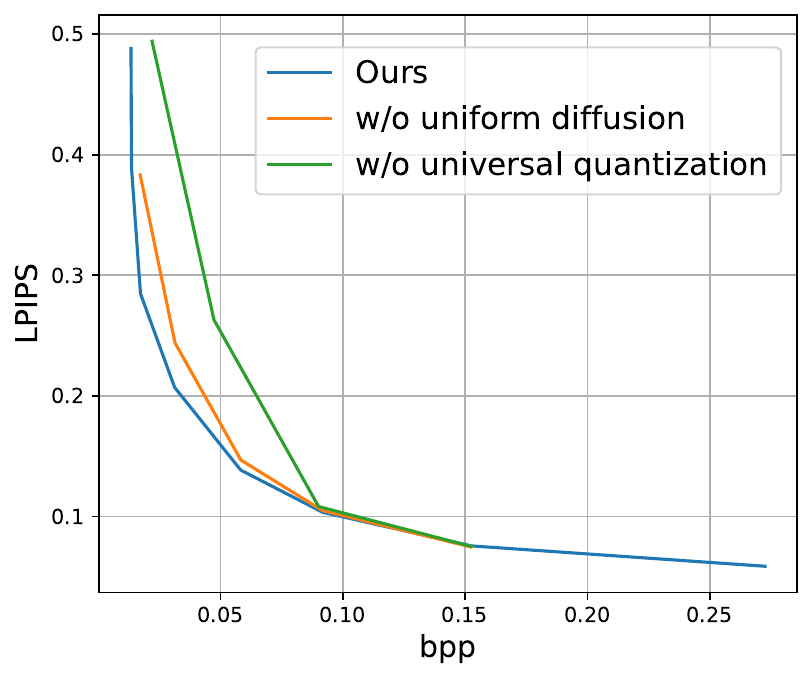}
    \caption{Rate-distortion results of our ablation study on the Kodak dataset.}
    \label{fig:ablation}
\end{figure} 

\def\picwidth{\linewidth}
\begin{figure*}[ht!]
    \begin{subfigure}{\textwidth}
        \centering
        \includegraphics[width=\picwidth]{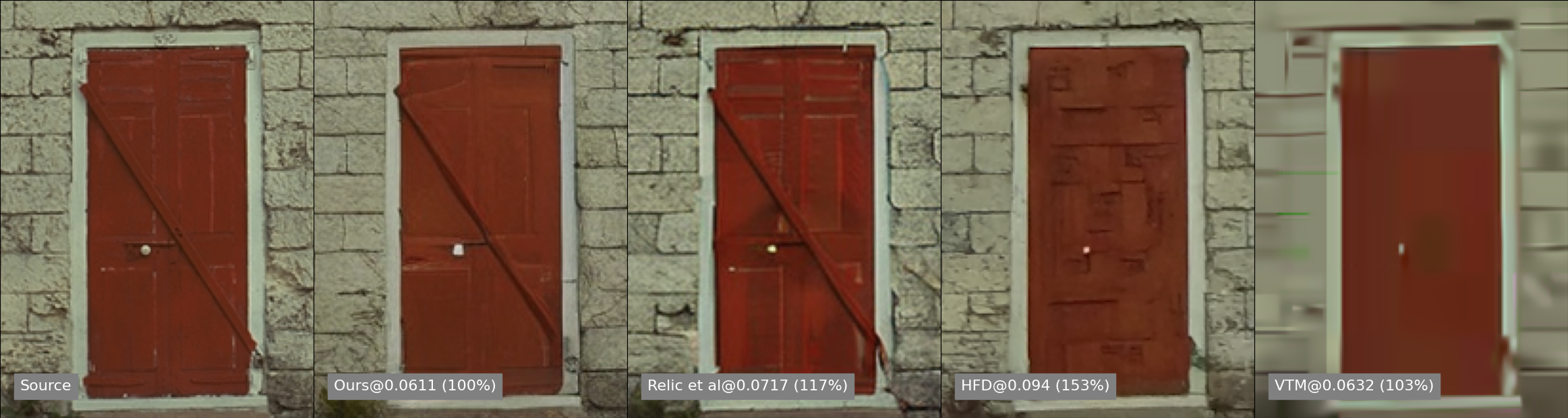}
    \end{subfigure}
    \begin{subfigure}{\textwidth}
        \centering
        \includegraphics[width=\picwidth]{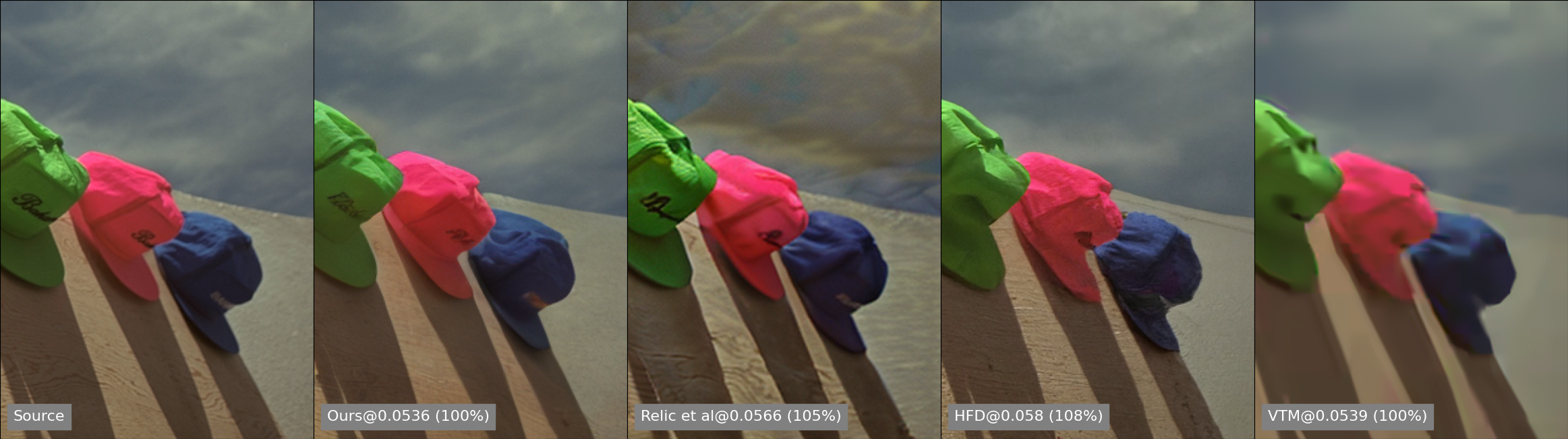}
    \end{subfigure}
    \begin{subfigure}{\textwidth}
        \centering
        \includegraphics[width=\picwidth]{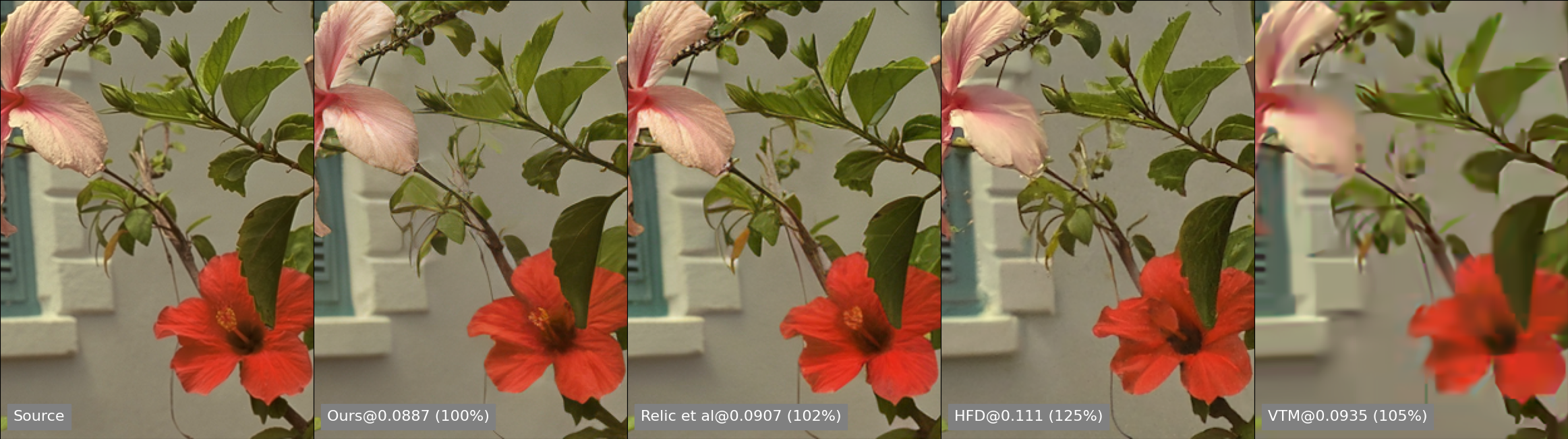}
    \end{subfigure}
    \begin{subfigure}{\textwidth}
        \centering
        \includegraphics[width=\picwidth]{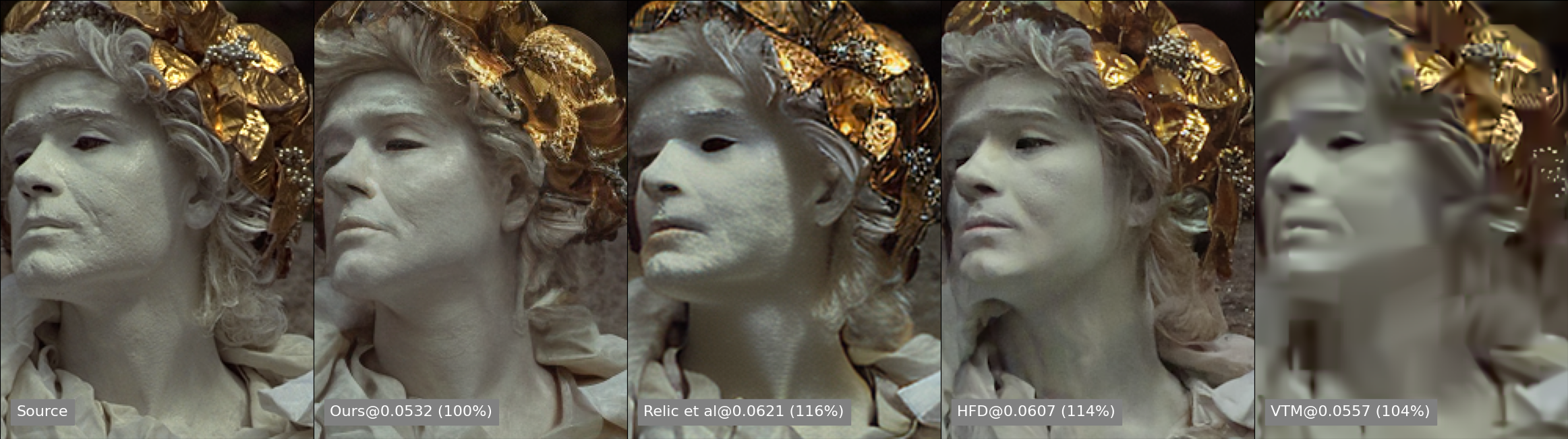}
    \end{subfigure}
    \caption{Qualitative comparison of our method to Relic et al.~\cite{relic2024Lossy}, HFD~\cite{hoogeboom2023HighFidelity}, and VTM~\cite{VTM192} on the Kodak dataset. Bitrates are also expressed as a percentage of our method. Best viewed digitally.}
\label{fig:more_visual_comparison}
\end{figure*}

Quantitative results are shown in Fig.~\ref{fig:rd}.
Our model achieves superior rate-realism performance on MS-COCO 30k below 0.1bpp and remains competitive with Relic~\etal at higher rates.
On the larger resolution images of the CLIC20 dataset, we achieve the best results among diffusion-based methods and similar performance as MS-ILLM over a wide range of bitrates.
When evaluating with LPIPS and MS-SSIM, we achieve the best performance below 0.08bpp.
Above this rate, our performance suffers when measured by MS-SSIM, consistent with the findings of Blau and Micheli~\cite{blau2019Rethinking} and other generative compression works~\cite{careil2023image, mentzer2020HighFidelitya}.

Qualitatively, as shown in Fig.~\ref{fig:more_visual_comparison}, our method consistently produces more detailed and accurate textures than HFD, especially noticeable in the red door and flowers.
The reconstructions of Relic~\etal significantly change the color compared to the ground truth, best shown in the sky and statue.
Our proposed method does not suffer from such artifacts and produces the most realistic reconstructions, maintaining a high accuracy to the original content while remaining free of any color shifts. 

As one of our goals is to improve DDIC at low bitrates, we examine the progression of image quality as bitrate decreases in Fig.~\ref{fig:bpp-progression}.
While Relic~\etal introduce gray color into the sky and over-saturate the wood grain or water, the reconstructions of our method maintain a high perceptual quality even into the extremely low bitrate range.

\subsubsection{Ablation}
To examine the impact of our proposed changes on overall performance we ablate the uniform diffusion model~(by denoising with a Gaussian diffusion model) and universal quantization~(by using hard quantization instead, \ie, \(\hat{\mathbf{z}}=\lfloor \sqrt{\alpha_t}\mathbf{y}\rceil\)).
The ablations are performed on the Kodak dataset and results shown in Fig.~\ref{fig:ablation}.
Our proposed changes improve the reconstruction quality of output images, especially at low bitrates.
This further shows the negative effects of leaving the three gaps unsolved, and reinforces that our proposed changes remove a barrier preventing DDIC-based methods from performing well at low bitrates.


\section{Conclusion}
\label{sec:conclusion}
While codecs following the DDIC paradigm benefit from increased computational efficiency and flexibilty of pretrained models, we show that there are three main barriers blocking effective compression to extremely low bitrates: the noise level gap, the noise type gap, and the discretization gap.
Our proposed quantization-based diffusion process and uniform noise diffusion model close these gaps, showing improved results at low bitrates while minimally affecting performance at higher rates.
We additionally are the first to show that uniform diffusion models work in the latent domain and on images of practical resolution, validating previous theoretical results.
Furthermore, we are the first to show that such models can be efficiently obtained by finetuning a Gaussian diffusion model on the desired distribution.
Potential future work includes addressing possible misgeneration of content, particularly in structural details at low birates.

\paragraph{Ethical Concerns.}
Generative compression, while powerful, raises ethical concerns due to the potential for mispredictions or misgeneration, where the model reconstructs details inaccurately or in a misleading way, potentially altering important information.
Future research must focus on addressing these risks, minimizing unintended alterations and maintaining the integrity of compressed data.


\newpage~\newpage

{
    \small
    \bibliographystyle{ieeenat_fullname}
    \bibliography{main}
}

\newpage
\clearpage
\maketitlesupplementary

\section{Additional Results}
\subsection{Qualitative Results}
\begin{figure*}
    \centering
    \includegraphics[width=0.95\textwidth]{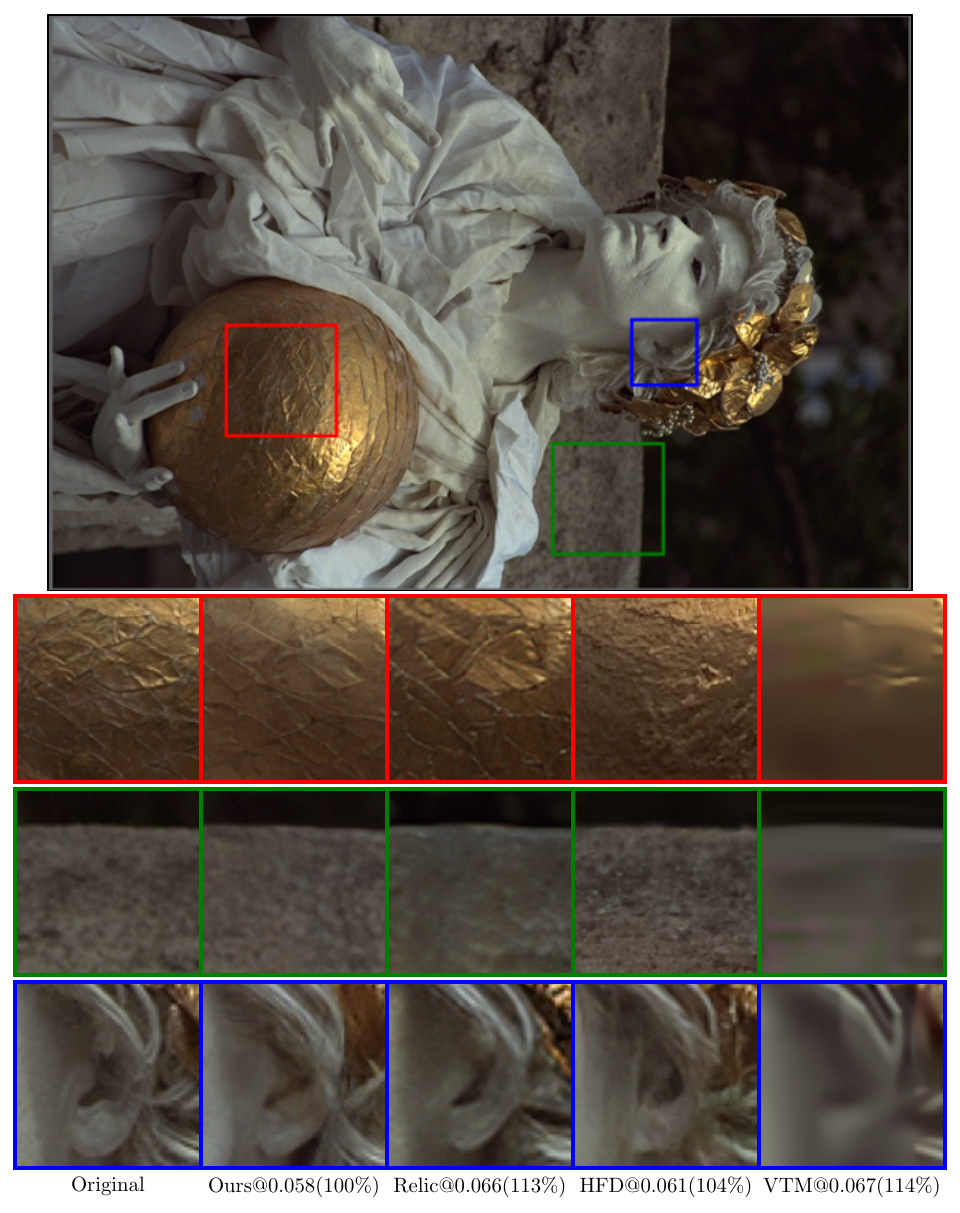}
    \caption{Additional visual comparisons on \emph{kodim17} between our method and Relic~\etal~\cite{relic2024Lossy}, HFD~\cite{hoogeboom2023HighFidelity}, and VTM~\cite{VTM192}. Images are labeled as [Method]@bpp and additionally shown as a percentage of our method. Here Relic~\etal flattens textures, and HFD incorrectly synthesizes content or oversharpens the image.}
\label{fig:addtl-crops-kodim17}
\end{figure*}

\begin{figure*}
    \centering
    \includegraphics[width=0.95\textwidth]{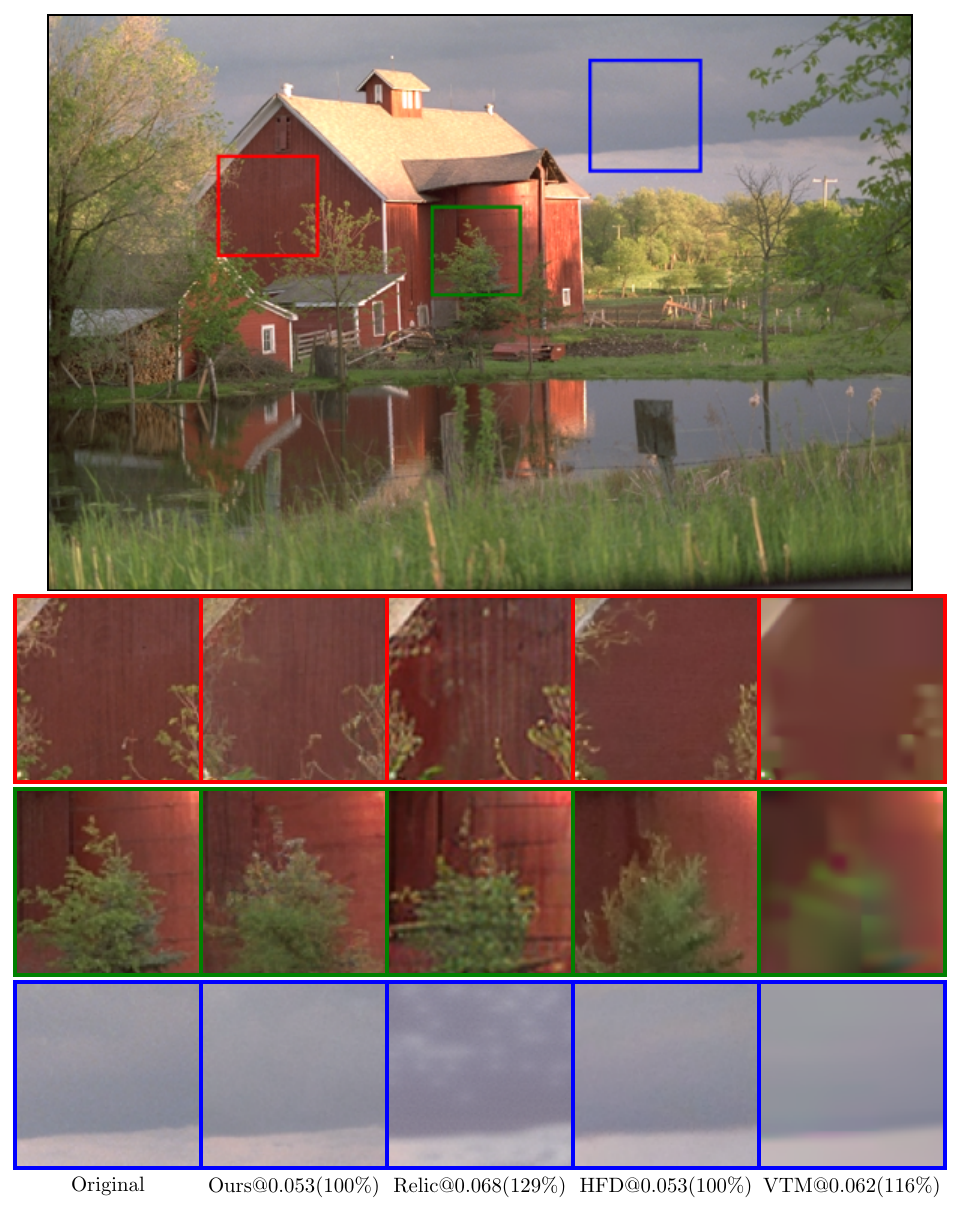}
    \caption{Additional visual comparisons on \emph{kodim22} between our method and Relic~\etal~\cite{relic2024Lossy}, HFD~\cite{hoogeboom2023HighFidelity}, and VTM~\cite{VTM192}. Images are labeled as [Method]@bpp and additionally shown as a percentage of our method. Here Relic~\etal introduces color artifacts and oversaturation, and HFD is not as detailed.}
\label{fig:addtl-crops-kodim22}
\end{figure*}

\begin{figure*}
    \centering
    \includegraphics[width=0.95\textwidth]{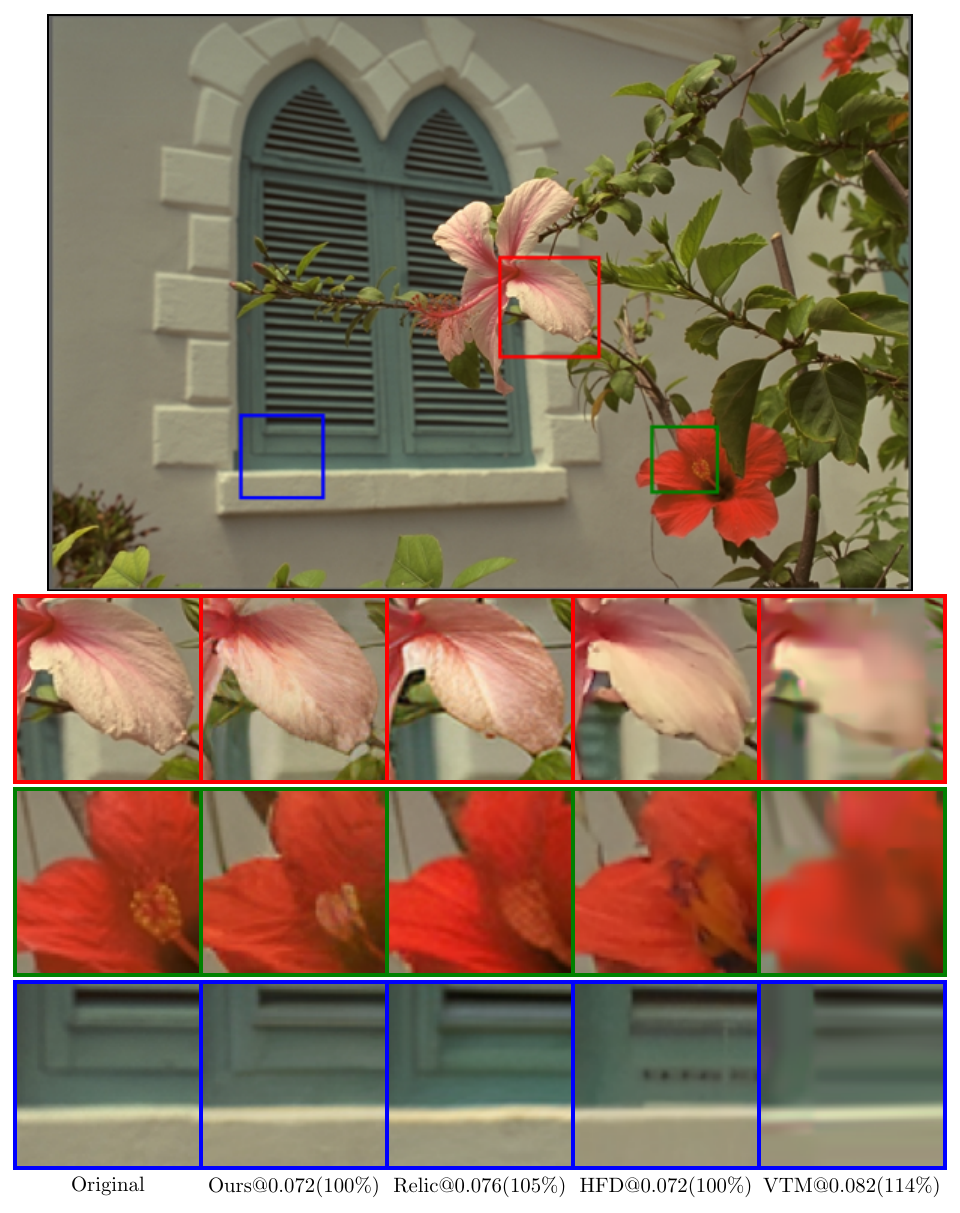}
    \caption{Additional visual comparisons on \emph{kodim07} between our method and Relic~\etal~\cite{relic2024Lossy}, HFD~\cite{hoogeboom2023HighFidelity}, and VTM~\cite{VTM192}. Images are labeled as [Method]@bpp and additionally shown as a percentage of our method. Here Relic~\etal and HFD do not correctly synthesize fine details or textures.}
\label{fig:addtl-crops-kodim07}
\end{figure*}

Additional visualizations are shown in Figs.~\ref{fig:addtl-crops-kodim17}-\ref{fig:addtl-crops-kodim07}.

\begin{figure*}
    \begin{subfigure}{\textwidth}
        \centering
        \includegraphics[width=\textwidth]{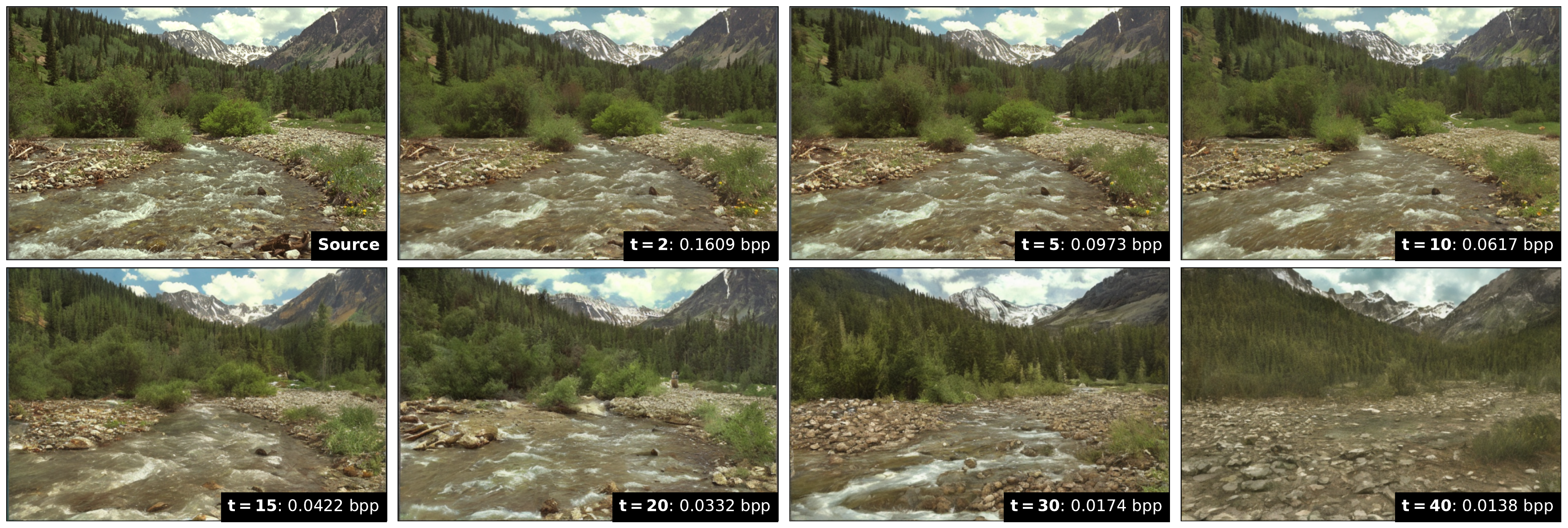}
    \end{subfigure}
    \par\medskip
    \begin{subfigure}{\textwidth}
        \centering
        \includegraphics[width=\textwidth]{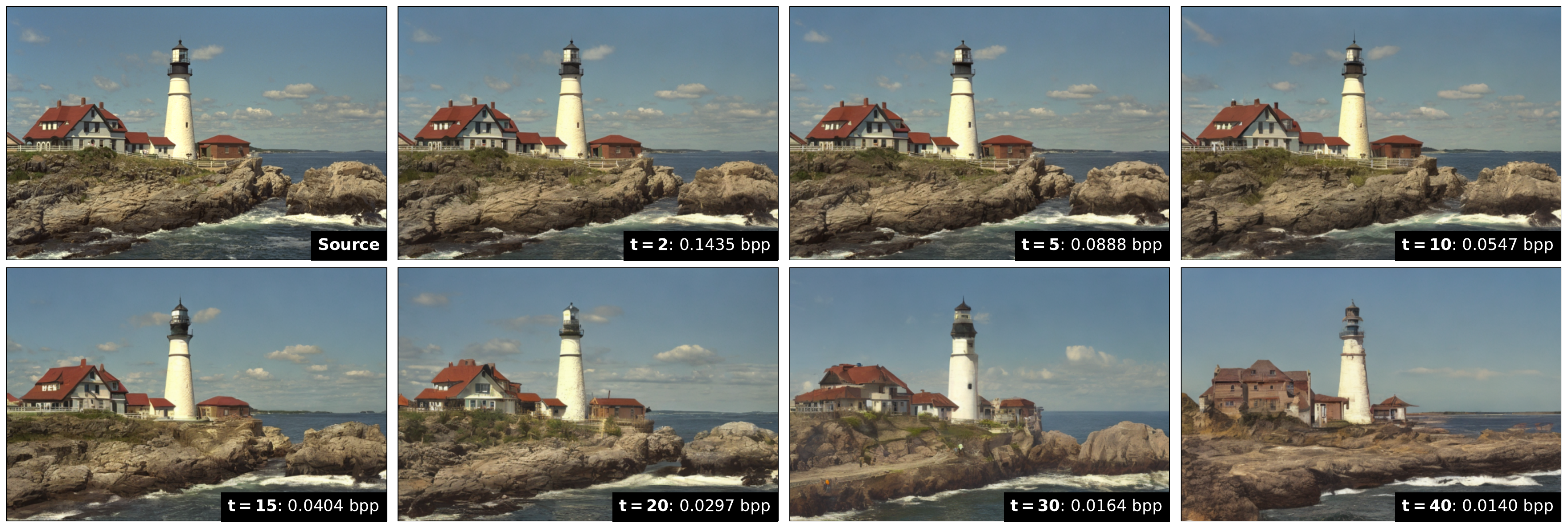}
    \end{subfigure}
    \par\medskip
    \begin{subfigure}{\textwidth}
        \centering
        \includegraphics[width=\textwidth]{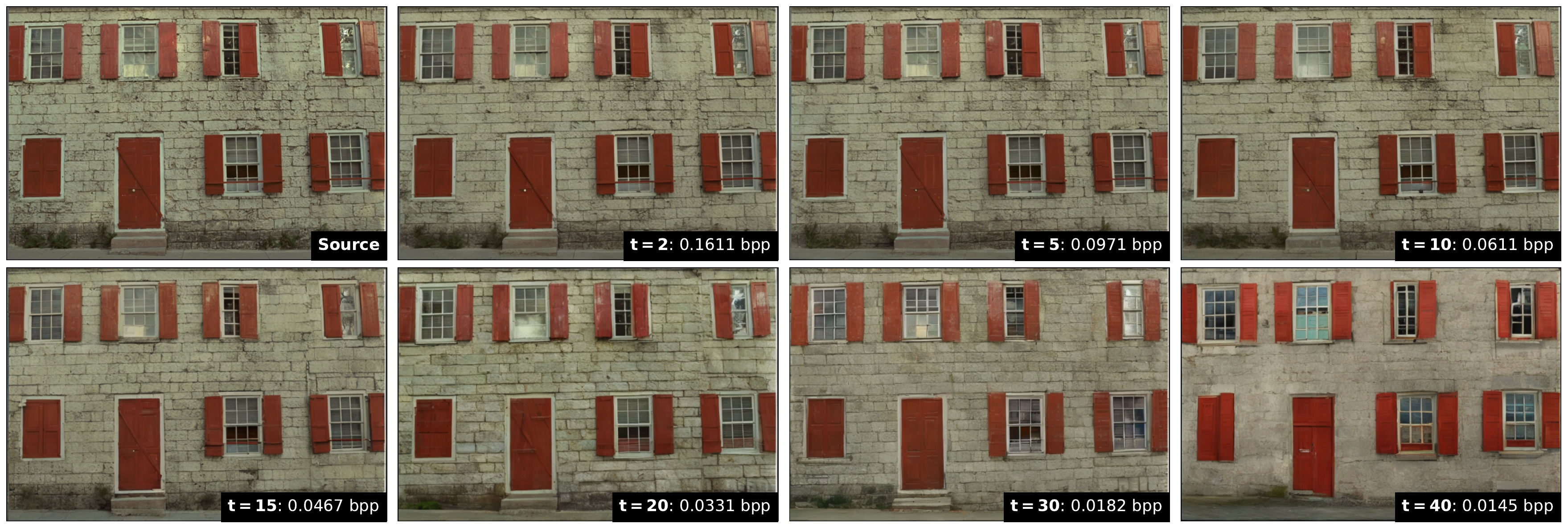}
    \end{subfigure}
    \caption{Visual example of image reconstructions produced over a range of bitrates. At high bitrate, our method produces reconstructions with high fidelity to the source image. As bitrate decreases, the images maintain a high realism and semantic alignment, tending towards differences in the spatial alignment of content. Best viewed digitally.}
\label{fig:bpp_progression}
\end{figure*}

Furthermore, to provide examples of our method's ability to consistently produce realistic and plausible images, we show examples of images compressed over a range of bitrates in Fig.~\ref{fig:bpp_progression}.
Here it can be seen that at high bitrates, the reconstructions are accurate at a pixel level to the source image.
As bitrate decreases, the images maintain high realism and semantic alignment and vary in low-level details~(\emph{e.g.}, bush, rock, or brick textures), rather than blurring artifacts traditionally seen in neural image compression.

\subsection{Quantitative Results}
\begin{figure}
    \centering
    \includegraphics[width=0.9\linewidth]{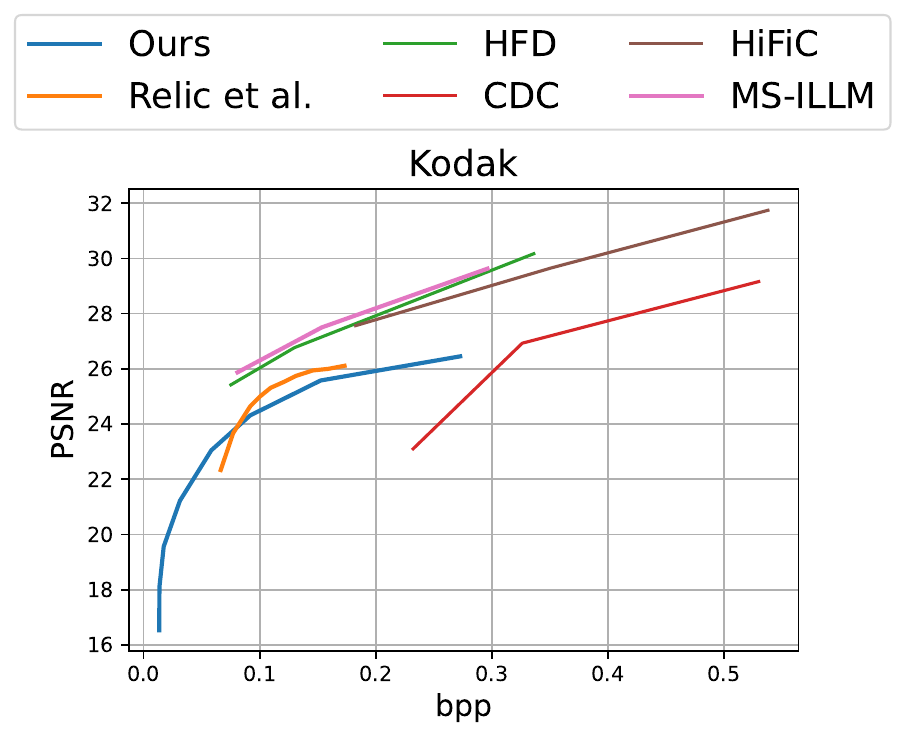}
    \caption{Rate-distortion comparison measured by PSNR on the Kodak dataset.}
    \label{fig:rd-psnr}
\end{figure}

In \cref{fig:rd-psnr}, we also include rate-distortion results measured in PSNR.
We emphasize that for generative compression tasks, PSNR does not accurately reflect the true visual quality of image reconstructions.
We include these results here solely for completeness.

\subsection{Ablation of Quantization Schedule}
While ablating the uniform diffusion model and universal quantization in our method is straightforward, ablating the quantization schedule is more arbitrary since we derive an objectively correct formulation.
However, it can be ablated as follows.

The quantization schedule controls two parameters in our proposed pipeline: the SNR of \(\hat{\mathbf{y}}_t\) and the number of denoising steps \(t\) to perform with the diffusion model at the receiver.
The former can also be quantified as a function of \(t\)~(as \(\Delta_t\) is dependent on \(t\); \cref{eq:uddq}); thus, we can consider two independent \(t\)s, one at the sender side~(\(t_s\)), and one at the receiver~(\(t_r\)).
With our quantization schedule we find a closed form solution for \(t_s=t_r\).
Therefore, to ablate the quantization schedule, we can manually sweep over a range of \(t_s\) and \(t_r\) and compute image quality metrics for all possible combinations.
Specifically, we choose all combinations of \(t_s, t_r \in \{1,2,5,10,20,30,40\}\) except where \(t_s =t_r\), as this corresponds to using our quantization schedule.

\begin{figure}
    \centering
    \includegraphics[width=0.9\linewidth]{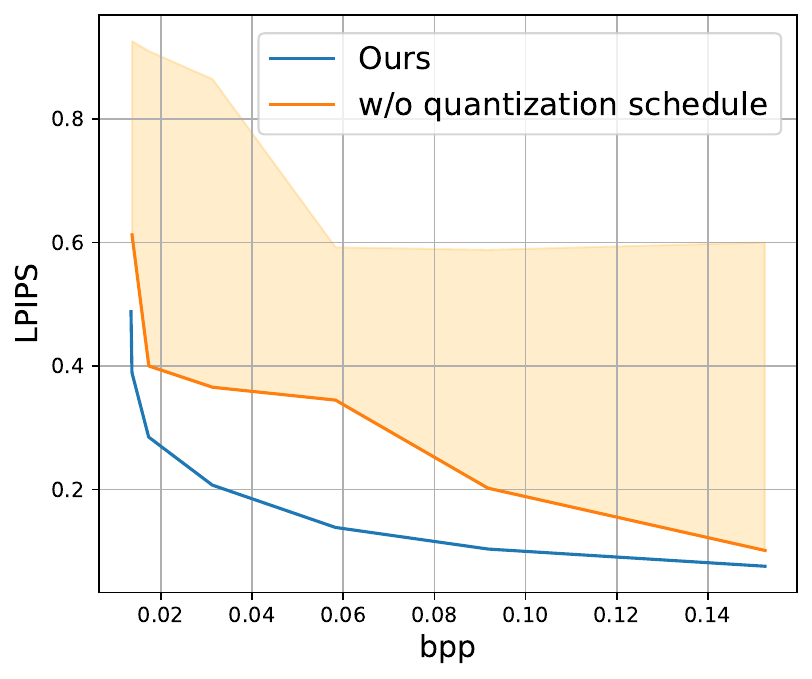}
    \caption{Ablation of our quantization schedule. The best performing combination of \(t_s\text{ and } t_r\) is shown, as well as the range of all tested combinations~(shaded).}
    \label{fig:ablation-quantschedule}
\end{figure}

Results are shown in \cref{fig:ablation-quantschedule}.
As in the other ablations, we can see the quantization schedule has a significant impact on image quality, particularly at low bitrates.

\subsection{Alternate Entropy Models}
\begin{figure*}
    \centering
    \includegraphics[width=0.9\linewidth]{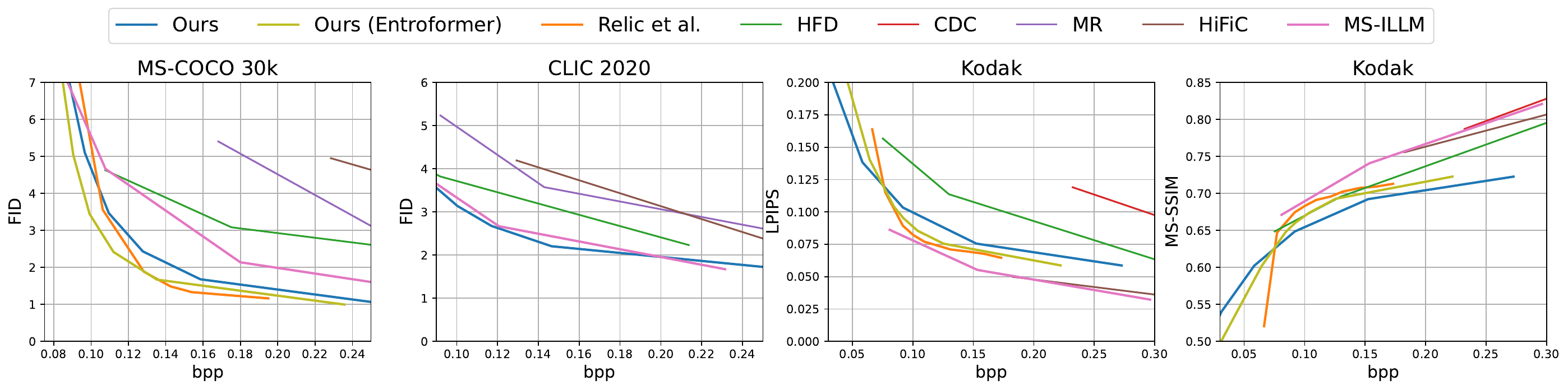}
    \caption{Additional rate-distortion comparisons including our method with an alternate entropy model.}
    \label{fig:supp-rd}
\end{figure*}

Our proposed method uses a relatively simple, yet standard mean-scale hyperprior entropy model~\cite{balle2017Endtoend,minnen2018Jointa}.
We also experimented with a newer, more complex entropy model (specifically Entroformer~\cite{yichen2022Entroformer}) used in the baseline method of Relic~\etal~\cite{relic2024Lossy}.
Quantitative results are shown in \cref{fig:supp-rd}.
It can be seen that the more powerful entropy model provides substantial rate-distortion performance gains in the higher bitrate range.
However, it suffers in the lower rates.

Given the additional complexity of including this entropy model in our proposal, as well as a significantly larger GPU memory requirement which prevents evaluation on high-resolution images, we elect to use the simpler mean-scale hyperprior entropy model in this work.

\section{Image Generation with Uniform Diffusion Models}

\begin{figure*}
    \centering
    \begin{subfigure}{0.95\textwidth}
        \centering
        \includegraphics[width=\linewidth]{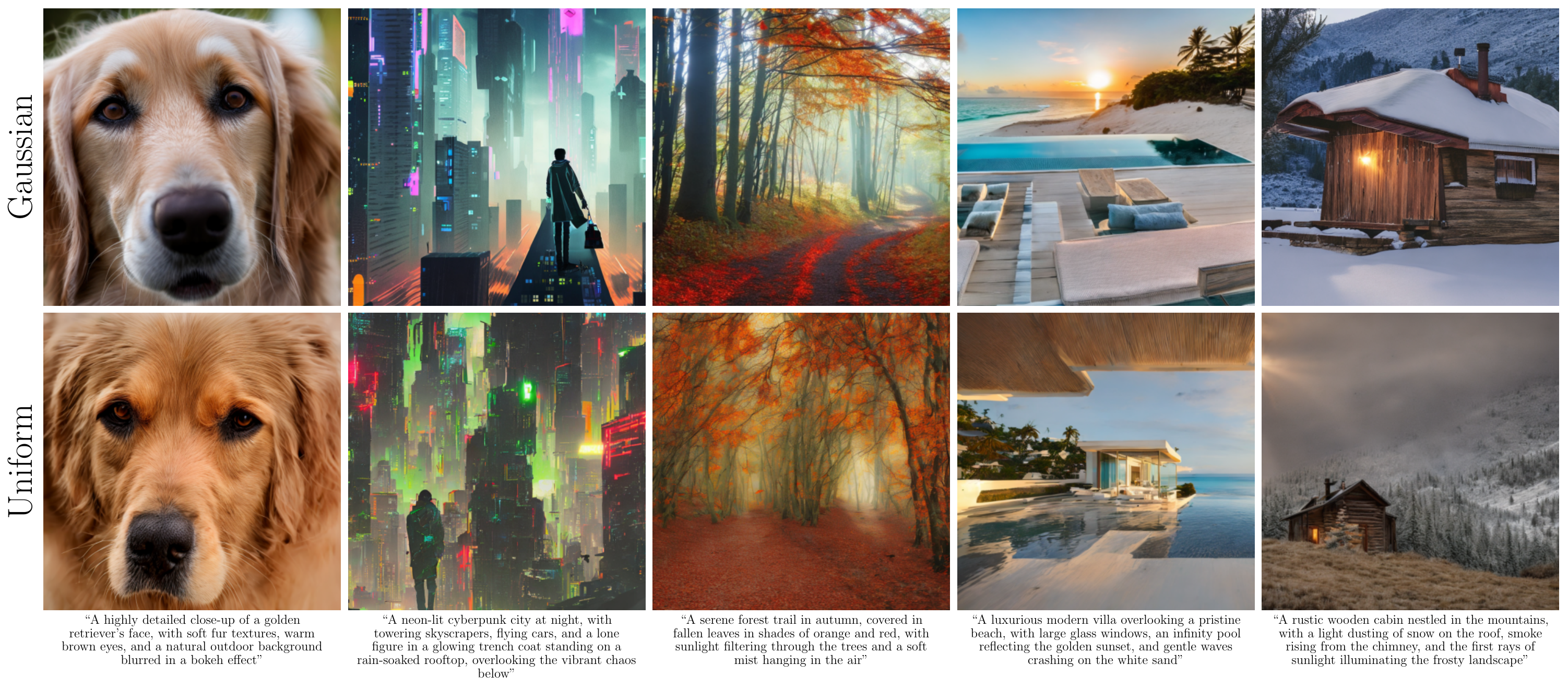}
    \end{subfigure}
    \begin{subfigure}{0.95\textwidth}
        \centering
        \includegraphics[width=\linewidth]{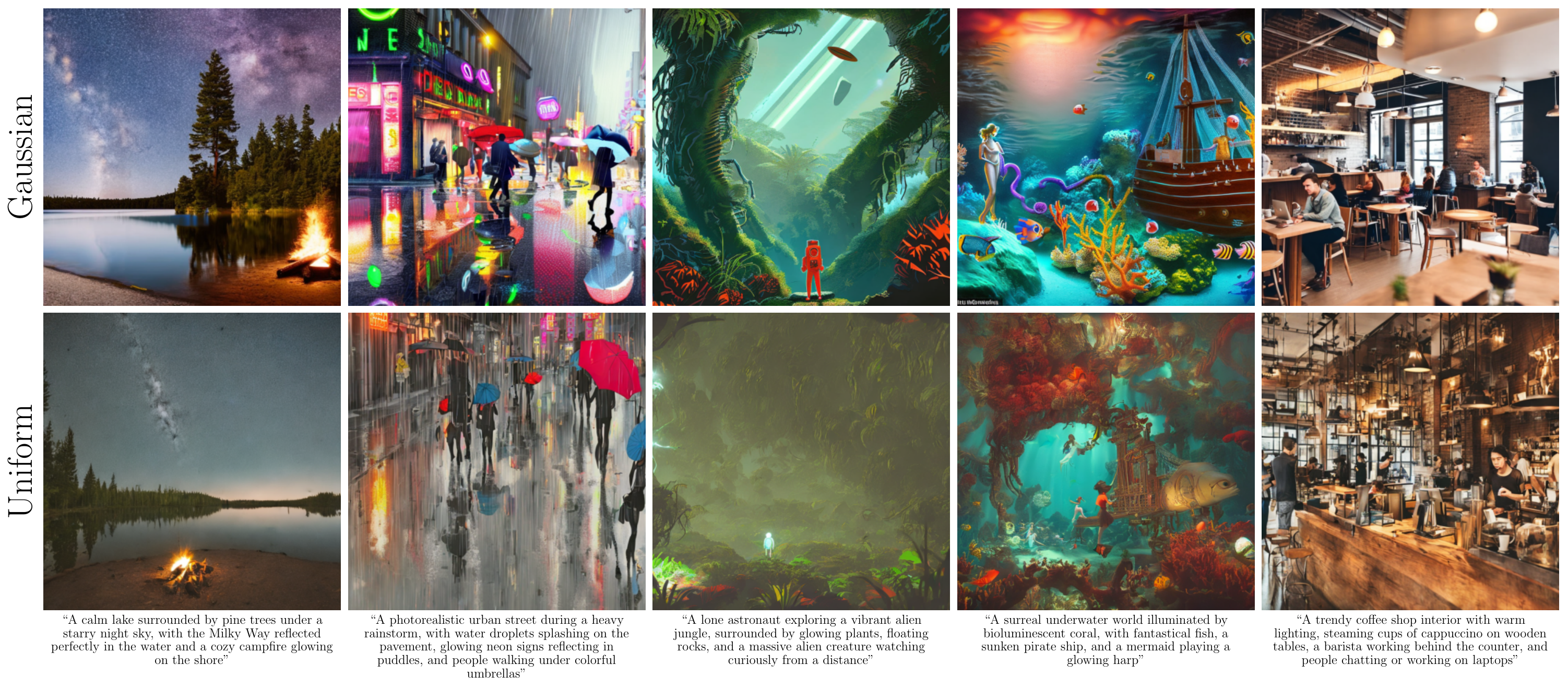}
    \end{subfigure}
    \begin{subfigure}{0.95\textwidth}
        \centering
        \includegraphics[width=\linewidth]{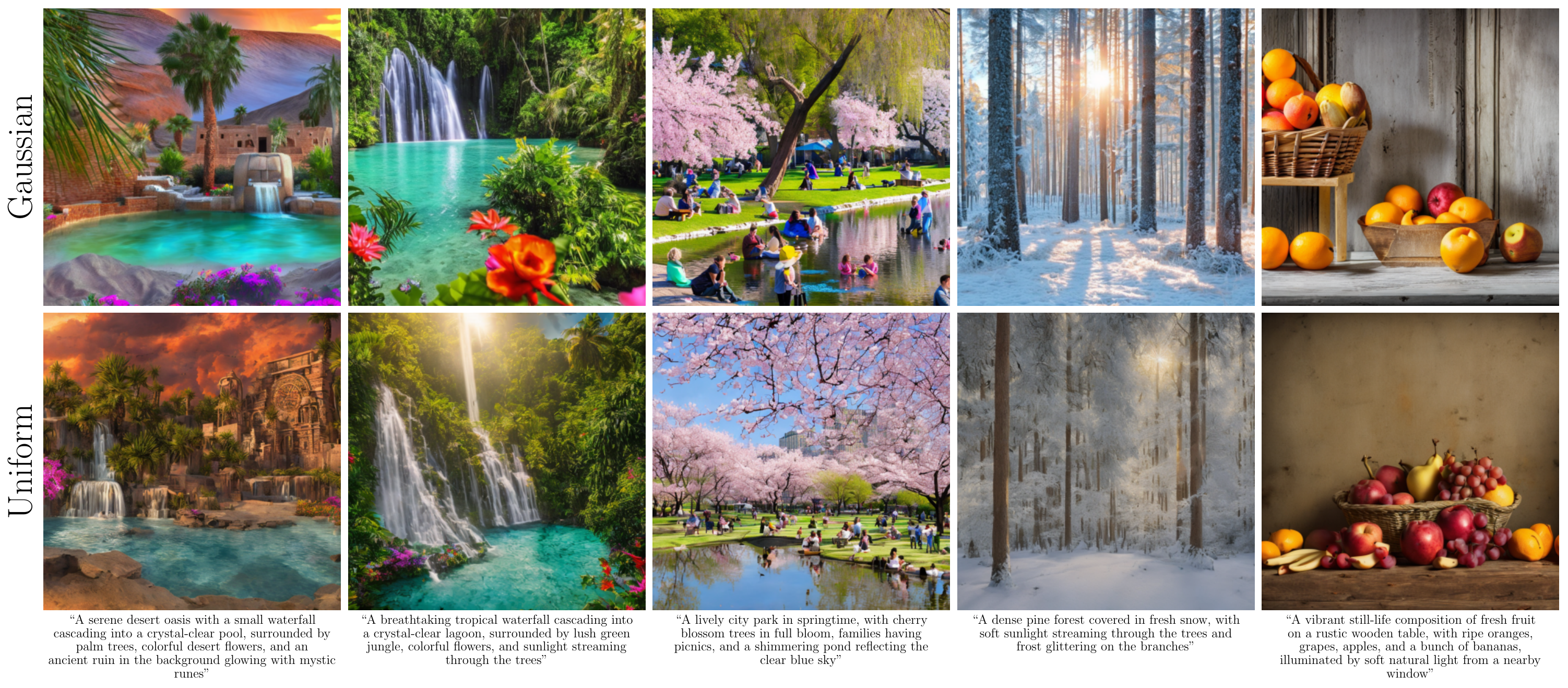}
    \end{subfigure}
    \caption{Image generation results between the standard Gaussian diffusion model versus our diffusion model finetuned for uniform noise. The prompt under each pair was arbitrarily generated using ChatGPT.}
    \label{fig:gauss-v-unif}
\end{figure*}

One concern when finetuning foundation diffusion models is catastrophic forgetting, where the model loses its ability to generate images.
To validate our uniform diffusion model retains its image generation capability, we compare generated images of our model with Stable Diffusion v2.1~(the starting weights for our finetuning) using the same prompt, shown in \cref{fig:gauss-v-unif}.
We perform text to image generation with DDIM sampling and generate images of \(768^2\) resolution.
The seed used to sample the initial noise is the same within each pair.
Prompts were arbitrarily generated by asking ChatGPT to write input prompts for Stable Diffusion with an emphasis on generating photorealistic images.

\section{Derivation of Quantization Schedule}
\label{supp:delta-derivation}

In this section we provide a derivation of our quantization schedule such that it matches the variance schedule of the original diffusion model~(\cref{eq:delta_t}). 
As described in \cref{sec:meat} and \cref{eq:quant_schedule}, this is done by matching the SNR of the partially noisy data \(\mathbf{y}_t\) of the original diffusion process with \(\hat{\mathbf{y}}_t\) of our proposed forward process.

The standard Gaussian diffusion process is defined as:

\begin{equation}
    \mathbf{y}_{t} = \sqrt{\alpha_{t}} \mathbf{y}_{0} + \epsilon, \quad \epsilon \sim \mathcal{N}(0, (1 - \alpha_{t})\mathbf{I}).
\end{equation}

In this context, the signal to noise ratio of \( \mathbf{y}_{t} \) is:
\begin{equation}
    \text{SNR}(\mathbf{y}_t) := \frac{\mathbb{E}[S^2]}{\text{Var}[N]} = \frac{\mathbb{E}[(\sqrt{\alpha_t}\mathbf{y}_0)^2]}{1-\alpha_t}.
\end{equation}

Using our proposed forward noising process in Eq.~\ref{eq:uddq} and deriving the SNR of \(\hat{\mathbf{y}}_t\):
\begin{align}
    \hat{\mathbf{y}}_t & = \lfloor \sqrt{\alpha_{t}}\mathbf{y}_0 - \mathbf{u} \rceil_{\Delta_t} + \mathbf{u}, \quad \mathbf{u} \sim \mathcal{U}[\nicefrac{-\Delta_t}{2}, \nicefrac{\Delta_t}{2}]\\
    & \stackrel{d}{=} \sqrt{\alpha_t}\mathbf{y}_0 + \mathbf{u}
\end{align}
\begin{align}
    \text{Var}[\mathbf{u}] & = \frac{1}{12}\left(\frac{\Delta_t}{2} - \frac{-\Delta_t}{2}\right)^2\\
    & = \frac{1}{12}\Delta_t^2
    \nonumber \\
    \therefore \quad \text{SNR}(\hat{\mathbf{y}}_t) & = \frac{\mathbb{E}[(\sqrt{\alpha_t}\mathbf{y}_0)^2]}{\frac{1}{12}\Delta_t^2}
\end{align}

Therefore, to match the SNR between Gaussian diffusion and our proposed method:

\begin{align}
    \text{SNR}(\hat{\mathbf{y}}_t) &= \text{SNR}(\mathbf{y}_t)\\
    \frac{\mathbb{E}[(\sqrt{\alpha_t}\mathbf{y}_0)^2]}{\frac{1}{12}\Delta_t^2} &= \frac{\mathbb{E}[(\sqrt{\alpha_t}\mathbf{y}_0)^2]}{1-\alpha_t}\\ 
    \frac{1}{12}\Delta_t^2 & = 1 - \alpha_t\\
    \Delta_t & = \sqrt{12(1-\alpha_t}).
\end{align}

\section{Entropy Coding with Variable Width Quantization Bins}
Entropy coding our latent \(\hat{\mathbf{z}}\) to bitstream requires a probability model over the set of symbols to be encoded – this is parameterized by the entropy model.
As discussed by Ball\'e~\etal~\cite{balle2017Endtoend, balle2018Variationala}, in order for the entropy model to better match the true distribution of transmitted symbols, the underlying density model of the entropy model is convolved with a box filter of unit width, which lends itself to easy computation by evaluating the cumulative distribution~(Eq.~(29) in \cite{balle2018Variationala}):
\begin{align}
    p_{\hat{\mathbf{z}}}(\hat{\mathbf{z}}) &= \Bigl ( p * \mathcal{U}(-\nicefrac{1}{2}, \nicefrac{1}{2})\Bigr ) (\hat{\mathbf{z}}) \\
    &= c(\hat{\mathbf{z}} + \nicefrac{1}{2}) - c(\hat{\mathbf{z}} - \nicefrac{1}{2})
\label{eq:prob_estimation_cdf}
\end{align}
where \(p\) is the underlying density model and \(c\) is the cumulative of \(p\).
Intuitively, to estimate the probability of some discretized value \(\hat{\mathbf{z}}_i\) we integrate the density model over the range of unquantized values that result in \(\hat{\mathbf{z}}_i\) after quantization.

However, an often overlooked fact is that the width of this box filter must equal the quantization bin width – using integer quantization requires a unit width box filter.
Therefore, when quantizing with arbitrary bin width, as done in this paper, we must implement an entropy model which estimates probabilities according to this width:
\begin{equation}
    p_{\hat{\mathbf{z}}}(\hat{\mathbf{z}}) = c(\hat{\mathbf{z}} + \nicefrac{\Delta}{2}) - c(\hat{\mathbf{z}} - \nicefrac{\Delta}{2})
\end{equation}

This does not lend itself to an intuitive implementation into current entropy modeling frameworks.
However, we find that this issue can be resolved, along with a simple implementation of non-integer quantization, by scaling \(\hat{\mathbf{z}}\) by \(\Delta_t\) before quantization and entropy coding and rescaling back upon decoding:

\begin{align}
    \hat{\mathbf{z}}= \left \lfloor \frac{\sqrt{\alpha_t}\mathbf{y}}{\Delta_t}-\mathbf{u}\right \rceil, \quad \hat{\mathbf{y}}_t=(\hat{\mathbf{z}} + \mathbf{u}) \cdot \Delta_t, \\
    \notag \text{where \;} \mathbf{u} \sim \mathcal{U}(-\nicefrac{1}{2}, \nicefrac{1}{2}) \text{\; and \;} \Delta_t=\sqrt{12(1-\alpha_t)}.
\end{align}
This is equivalent to \cref{eq:uddq} but allows the standard \cref{eq:prob_estimation_cdf} to be used for probability estimation.
This requires \(\Delta_t\) to be known by the reciever, however incurs no rate penalty as \(\Delta_t\) is derived from \(t\) which is already transmitted as side information.

\section{Reproduction of Baselines}
In the following paragraphs we detail specifics on how image reconstructiond and quantitative metrics were obtained for each baseline.

\textbf{Relic~\etal.}
Reconstructions and quantitative metrics were obtained through personal communication.

\textbf{HFD.}
Reconstructions were obtained through personal communication, also available at \texttt{\url{http://theis.io/hifidiff//}}. Quantitative metrics are used as reported in their paper.

\textbf{MR.}
Quantitative metrics are used as reported in their paper.

\textbf{HiFiC.}
Pretrained models are available in Tensorflow Compression \texttt{tfci} as \texttt{hific-hi}, \texttt{hific-mi}, and \texttt{hific-lo}.

\textbf{CDC.}
We produce images on our evaluation datasets using the official code release at \texttt{\url{https://github.com/buggyyang/CDC_compression}} (commit \texttt{742de7f}).
We take the epsilon parameterized models trained with \(L_1\) loss and axuillary LPIPS perceptual loss with weighting parameter \(\rho=0.9\).
During inference, we use 1000 sampling steps.

\textbf{VTM}
We use VTM 19.2, available at \texttt{\url{https://vcgit.hhi.fraunhofer.de/jvet/VVCSoftware_VTM/-/releases/VTM-19.2}} and run with the following commands:
\begin{verbatim}
# Encode
    EncoderAppStatic
    -c encoder_intra_vtm.cfg
    -i $INPUT
    -q $QP
    -o /dev/null
    -b $OUTPUT
    -wdt $WIDTH
    -hgt $HEIGHT
    -fr 1
    -f 1
    --InputChromaFormat=444
    --InputBitDepth=8
    --ConformanceWindowMode=1
    --InputColourSpaceConvert=RGBtoGBR
    --SNRInternalColourSpace=1
    --OutputInternalColourScace=0

# Decode
    DecoderAppStatic
    -b $OUTPUT
    -o $RECON
    -d 8
    --OutputColourSpaceConvert=GBRtoRGB
\end{verbatim}

\end{document}